\documentclass[prl,superscriptaddress,twocolumn,nofootinbib,showpacs,preprintnumbers,floatfix]{revtex4}

\usepackage{mathrsfs}
\usepackage{amsfonts,amssymb,amsmath}
\usepackage{graphicx,epsfig}
\usepackage{multirow}
\usepackage{verbatim}
\usepackage{color}
\definecolor{AliceBlue}{rgb}{0.94,0.97,1.00}
\definecolor{AntiqueWhite1}{rgb}{1.00,0.94,0.86}
\definecolor{AntiqueWhite2}{rgb}{0.93,0.87,0.80}
\definecolor{AntiqueWhite3}{rgb}{0.80,0.75,0.69}
\definecolor{AntiqueWhite4}{rgb}{0.55,0.51,0.47}
\definecolor{AntiqueWhite}{rgb}{0.98,0.92,0.84}
\definecolor{BlanchedAlmond}{rgb}{1.00,0.92,0.80}
\definecolor{BlueViolet}{rgb}{0.54,0.17,0.89}
\definecolor{CadetBlue1}{rgb}{0.60,0.96,1.00}
\definecolor{CadetBlue2}{rgb}{0.56,0.90,0.93}
\definecolor{CadetBlue3}{rgb}{0.48,0.77,0.80}
\definecolor{CadetBlue4}{rgb}{0.33,0.53,0.55}
\definecolor{CadetBlue}{rgb}{0.37,0.62,0.63}
\definecolor{CornflowerBlue}{rgb}{0.39,0.58,0.93}
\definecolor{DarkBlue}{rgb}{0.00,0.00,0.55}
\definecolor{DarkCyan}{rgb}{0.00,0.55,0.55}
\definecolor{DarkGoldenrod1}{rgb}{1.00,0.73,0.06}
\definecolor{DarkGoldenrod2}{rgb}{0.93,0.68,0.05}
\definecolor{DarkGoldenrod3}{rgb}{0.80,0.58,0.05}
\definecolor{DarkGoldenrod4}{rgb}{0.55,0.40,0.03}
\definecolor{DarkGoldenrod}{rgb}{0.72,0.53,0.04}
\definecolor{DarkGray}{rgb}{0.66,0.66,0.66}
\definecolor{DarkGreen}{rgb}{0.00,0.39,0.00}
\definecolor{DarkGrey}{rgb}{0.66,0.66,0.66}
\definecolor{DarkKhaki}{rgb}{0.74,0.72,0.42}
\definecolor{DarkMagenta}{rgb}{0.55,0.00,0.55}
\definecolor{DarkOliveGreen1}{rgb}{0.79,1.00,0.44}
\definecolor{DarkOliveGreen2}{rgb}{0.74,0.93,0.41}
\definecolor{DarkOliveGreen3}{rgb}{0.64,0.80,0.35}
\definecolor{DarkOliveGreen4}{rgb}{0.43,0.55,0.24}
\definecolor{DarkOliveGreen}{rgb}{0.33,0.42,0.18}
\definecolor{DarkOrange1}{rgb}{1.00,0.50,0.00}
\definecolor{DarkOrange2}{rgb}{0.93,0.46,0.00}
\definecolor{DarkOrange3}{rgb}{0.80,0.40,0.00}
\definecolor{DarkOrange4}{rgb}{0.55,0.27,0.00}
\definecolor{DarkOrange}{rgb}{1.00,0.55,0.00}
\definecolor{DarkOrchid1}{rgb}{0.75,0.24,1.00}
\definecolor{DarkOrchid2}{rgb}{0.70,0.23,0.93}
\definecolor{DarkOrchid3}{rgb}{0.60,0.20,0.80}
\definecolor{DarkOrchid4}{rgb}{0.41,0.13,0.55}
\definecolor{DarkOrchid}{rgb}{0.60,0.20,0.80}
\definecolor{DarkRed}{rgb}{0.55,0.00,0.00}
\definecolor{DarkSalmon}{rgb}{0.91,0.59,0.48}
\definecolor{DarkSeaGreen1}{rgb}{0.76,1.00,0.76}
\definecolor{DarkSeaGreen2}{rgb}{0.71,0.93,0.71}
\definecolor{DarkSeaGreen3}{rgb}{0.61,0.80,0.61}
\definecolor{DarkSeaGreen4}{rgb}{0.41,0.55,0.41}
\definecolor{DarkSeaGreen}{rgb}{0.56,0.74,0.56}
\definecolor{DarkSlateBlue}{rgb}{0.28,0.24,0.55}
\definecolor{DarkSlateGray1}{rgb}{0.59,1.00,1.00}
\definecolor{DarkSlateGray2}{rgb}{0.55,0.93,0.93}
\definecolor{DarkSlateGray3}{rgb}{0.47,0.80,0.80}
\definecolor{DarkSlateGray4}{rgb}{0.32,0.55,0.55}
\definecolor{DarkSlateGray}{rgb}{0.18,0.31,0.31}
\definecolor{DarkSlateGrey}{rgb}{0.18,0.31,0.31}
\definecolor{DarkTurquoise}{rgb}{0.00,0.81,0.82}
\definecolor{DarkViolet}{rgb}{0.58,0.00,0.83}
\definecolor{DeepPink1}{rgb}{1.00,0.08,0.58}
\definecolor{DeepPink2}{rgb}{0.93,0.07,0.54}
\definecolor{DeepPink3}{rgb}{0.80,0.06,0.46}
\definecolor{DeepPink4}{rgb}{0.55,0.04,0.31}
\definecolor{DeepPink}{rgb}{1.00,0.08,0.58}
\definecolor{DeepSkyBlue1}{rgb}{0.00,0.75,1.00}
\definecolor{DeepSkyBlue2}{rgb}{0.00,0.70,0.93}
\definecolor{DeepSkyBlue3}{rgb}{0.00,0.60,0.80}
\definecolor{DeepSkyBlue4}{rgb}{0.00,0.41,0.55}
\definecolor{DeepSkyBlue}{rgb}{0.00,0.75,1.00}
\definecolor{DimGray}{rgb}{0.41,0.41,0.41}
\definecolor{DimGrey}{rgb}{0.41,0.41,0.41}
\definecolor{DodgerBlue1}{rgb}{0.12,0.56,1.00}
\definecolor{DodgerBlue2}{rgb}{0.11,0.53,0.93}
\definecolor{DodgerBlue3}{rgb}{0.09,0.45,0.80}
\definecolor{DodgerBlue4}{rgb}{0.06,0.31,0.55}
\definecolor{DodgerBlue}{rgb}{0.12,0.56,1.00}
\definecolor{FloralWhite}{rgb}{1.00,0.98,0.94}
\definecolor{ForestGreen}{rgb}{0.13,0.55,0.13}
\definecolor{GhostWhite}{rgb}{0.97,0.97,1.00}
\definecolor{GreenYellow}{rgb}{0.68,1.00,0.18}
\definecolor{HotPink1}{rgb}{1.00,0.43,0.71}
\definecolor{HotPink2}{rgb}{0.93,0.42,0.65}
\definecolor{HotPink3}{rgb}{0.80,0.38,0.56}
\definecolor{HotPink4}{rgb}{0.55,0.23,0.38}
\definecolor{HotPink}{rgb}{1.00,0.41,0.71}
\definecolor{IndianRed1}{rgb}{1.00,0.42,0.42}
\definecolor{IndianRed2}{rgb}{0.93,0.39,0.39}
\definecolor{IndianRed3}{rgb}{0.80,0.33,0.33}
\definecolor{IndianRed4}{rgb}{0.55,0.23,0.23}
\definecolor{IndianRed}{rgb}{0.80,0.36,0.36}
\definecolor{LavenderBlush1}{rgb}{1.00,0.94,0.96}
\definecolor{LavenderBlush2}{rgb}{0.93,0.88,0.90}
\definecolor{LavenderBlush3}{rgb}{0.80,0.76,0.77}
\definecolor{LavenderBlush4}{rgb}{0.55,0.51,0.53}
\definecolor{LavenderBlush}{rgb}{1.00,0.94,0.96}
\definecolor{LawnGreen}{rgb}{0.49,0.99,0.00}
\definecolor{LemonChiffon1}{rgb}{1.00,0.98,0.80}
\definecolor{LemonChiffon2}{rgb}{0.93,0.91,0.75}
\definecolor{LemonChiffon3}{rgb}{0.80,0.79,0.65}
\definecolor{LemonChiffon4}{rgb}{0.55,0.54,0.44}
\definecolor{LemonChiffon}{rgb}{1.00,0.98,0.80}
\definecolor{LightBlue1}{rgb}{0.75,0.94,1.00}
\definecolor{LightBlue2}{rgb}{0.70,0.87,0.93}
\definecolor{LightBlue3}{rgb}{0.60,0.75,0.80}
\definecolor{LightBlue4}{rgb}{0.41,0.51,0.55}
\definecolor{LightBlue}{rgb}{0.68,0.85,0.90}
\definecolor{LightCoral}{rgb}{0.94,0.50,0.50}
\definecolor{LightCyan1}{rgb}{0.88,1.00,1.00}
\definecolor{LightCyan2}{rgb}{0.82,0.93,0.93}
\definecolor{LightCyan3}{rgb}{0.71,0.80,0.80}
\definecolor{LightCyan4}{rgb}{0.48,0.55,0.55}
\definecolor{LightCyan}{rgb}{0.88,1.00,1.00}
\definecolor{LightGoldenrod1}{rgb}{1.00,0.93,0.55}
\definecolor{LightGoldenrod2}{rgb}{0.93,0.86,0.51}
\definecolor{LightGoldenrod3}{rgb}{0.80,0.75,0.44}
\definecolor{LightGoldenrod4}{rgb}{0.55,0.51,0.30}
\definecolor{LightGoldenrodYellow}{rgb}{0.98,0.98,0.82}
\definecolor{LightGoldenrod}{rgb}{0.93,0.87,0.51}
\definecolor{LightGray}{rgb}{0.83,0.83,0.83}
\definecolor{LightGreen}{rgb}{0.56,0.93,0.56}
\definecolor{LightGrey}{rgb}{0.83,0.83,0.83}
\definecolor{LightPink1}{rgb}{1.00,0.68,0.73}
\definecolor{LightPink2}{rgb}{0.93,0.64,0.68}
\definecolor{LightPink3}{rgb}{0.80,0.55,0.58}
\definecolor{LightPink4}{rgb}{0.55,0.37,0.40}
\definecolor{LightPink}{rgb}{1.00,0.71,0.76}
\definecolor{LightSalmon1}{rgb}{1.00,0.63,0.48}
\definecolor{LightSalmon2}{rgb}{0.93,0.58,0.45}
\definecolor{LightSalmon3}{rgb}{0.80,0.51,0.38}
\definecolor{LightSalmon4}{rgb}{0.55,0.34,0.26}
\definecolor{LightSalmon}{rgb}{1.00,0.63,0.48}
\definecolor{LightSeaGreen}{rgb}{0.13,0.70,0.67}
\definecolor{LightSkyBlue1}{rgb}{0.69,0.89,1.00}
\definecolor{LightSkyBlue2}{rgb}{0.64,0.83,0.93}
\definecolor{LightSkyBlue3}{rgb}{0.55,0.71,0.80}
\definecolor{LightSkyBlue4}{rgb}{0.38,0.48,0.55}
\definecolor{LightSkyBlue}{rgb}{0.53,0.81,0.98}
\definecolor{LightSlateBlue}{rgb}{0.52,0.44,1.00}
\definecolor{LightSlateGray}{rgb}{0.47,0.53,0.60}
\definecolor{LightSlateGrey}{rgb}{0.47,0.53,0.60}
\definecolor{LightSteelBlue1}{rgb}{0.79,0.88,1.00}
\definecolor{LightSteelBlue2}{rgb}{0.74,0.82,0.93}
\definecolor{LightSteelBlue3}{rgb}{0.64,0.71,0.80}
\definecolor{LightSteelBlue4}{rgb}{0.43,0.48,0.55}
\definecolor{LightSteelBlue}{rgb}{0.69,0.77,0.87}
\definecolor{LightYellow1}{rgb}{1.00,1.00,0.88}
\definecolor{LightYellow2}{rgb}{0.93,0.93,0.82}
\definecolor{LightYellow3}{rgb}{0.80,0.80,0.71}
\definecolor{LightYellow4}{rgb}{0.55,0.55,0.48}
\definecolor{LightYellow}{rgb}{1.00,1.00,0.88}
\definecolor{LimeGreen}{rgb}{0.20,0.80,0.20}
\definecolor{MediumAquamarine}{rgb}{0.40,0.80,0.67}
\definecolor{MediumBlue}{rgb}{0.00,0.00,0.80}
\definecolor{MediumOrchid1}{rgb}{0.88,0.40,1.00}
\definecolor{MediumOrchid2}{rgb}{0.82,0.37,0.93}
\definecolor{MediumOrchid3}{rgb}{0.71,0.32,0.80}
\definecolor{MediumOrchid4}{rgb}{0.48,0.22,0.55}
\definecolor{MediumOrchid}{rgb}{0.73,0.33,0.83}
\definecolor{MediumPurple1}{rgb}{0.67,0.51,1.00}
\definecolor{MediumPurple2}{rgb}{0.62,0.47,0.93}
\definecolor{MediumPurple3}{rgb}{0.54,0.41,0.80}
\definecolor{MediumPurple4}{rgb}{0.36,0.28,0.55}
\definecolor{MediumPurple}{rgb}{0.58,0.44,0.86}
\definecolor{MediumSeaGreen}{rgb}{0.24,0.70,0.44}
\definecolor{MediumSlateBlue}{rgb}{0.48,0.41,0.93}
\definecolor{MediumSpringGreen}{rgb}{0.00,0.98,0.60}
\definecolor{MediumTurquoise}{rgb}{0.28,0.82,0.80}
\definecolor{MediumVioletRed}{rgb}{0.78,0.08,0.52}
\definecolor{MidnightBlue}{rgb}{0.10,0.10,0.44}
\definecolor{MintCream}{rgb}{0.96,1.00,0.98}
\definecolor{MistyRose1}{rgb}{1.00,0.89,0.88}
\definecolor{MistyRose2}{rgb}{0.93,0.84,0.82}
\definecolor{MistyRose3}{rgb}{0.80,0.72,0.71}
\definecolor{MistyRose4}{rgb}{0.55,0.49,0.48}
\definecolor{MistyRose}{rgb}{1.00,0.89,0.88}
\definecolor{NavajoWhite1}{rgb}{1.00,0.87,0.68}
\definecolor{NavajoWhite2}{rgb}{0.93,0.81,0.63}
\definecolor{NavajoWhite3}{rgb}{0.80,0.70,0.55}
\definecolor{NavajoWhite4}{rgb}{0.55,0.47,0.37}
\definecolor{NavajoWhite}{rgb}{1.00,0.87,0.68}
\definecolor{NavyBlue}{rgb}{0.00,0.00,0.50}
\definecolor{OldLace}{rgb}{0.99,0.96,0.90}
\definecolor{OliveDrab1}{rgb}{0.75,1.00,0.24}
\definecolor{OliveDrab2}{rgb}{0.70,0.93,0.23}
\definecolor{OliveDrab3}{rgb}{0.60,0.80,0.20}
\definecolor{OliveDrab4}{rgb}{0.41,0.55,0.13}
\definecolor{OliveDrab}{rgb}{0.42,0.56,0.14}
\definecolor{OrangeRed1}{rgb}{1.00,0.27,0.00}
\definecolor{OrangeRed2}{rgb}{0.93,0.25,0.00}
\definecolor{OrangeRed3}{rgb}{0.80,0.22,0.00}
\definecolor{OrangeRed4}{rgb}{0.55,0.15,0.00}
\definecolor{OrangeRed}{rgb}{1.00,0.27,0.00}
\definecolor{PaleGoldenrod}{rgb}{0.93,0.91,0.67}
\definecolor{PaleGreen1}{rgb}{0.60,1.00,0.60}
\definecolor{PaleGreen2}{rgb}{0.56,0.93,0.56}
\definecolor{PaleGreen3}{rgb}{0.49,0.80,0.49}
\definecolor{PaleGreen4}{rgb}{0.33,0.55,0.33}
\definecolor{PaleGreen}{rgb}{0.60,0.98,0.60}
\definecolor{PaleTurquoise1}{rgb}{0.73,1.00,1.00}
\definecolor{PaleTurquoise2}{rgb}{0.68,0.93,0.93}
\definecolor{PaleTurquoise3}{rgb}{0.59,0.80,0.80}
\definecolor{PaleTurquoise4}{rgb}{0.40,0.55,0.55}
\definecolor{PaleTurquoise}{rgb}{0.69,0.93,0.93}
\definecolor{PaleVioletRed1}{rgb}{1.00,0.51,0.67}
\definecolor{PaleVioletRed2}{rgb}{0.93,0.47,0.62}
\definecolor{PaleVioletRed3}{rgb}{0.80,0.41,0.54}
\definecolor{PaleVioletRed4}{rgb}{0.55,0.28,0.36}
\definecolor{PaleVioletRed}{rgb}{0.86,0.44,0.58}
\definecolor{PapayaWhip}{rgb}{1.00,0.94,0.84}
\definecolor{PeachPuff1}{rgb}{1.00,0.85,0.73}
\definecolor{PeachPuff2}{rgb}{0.93,0.80,0.68}
\definecolor{PeachPuff3}{rgb}{0.80,0.69,0.58}
\definecolor{PeachPuff4}{rgb}{0.55,0.47,0.40}
\definecolor{PeachPuff}{rgb}{1.00,0.85,0.73}
\definecolor{PowderBlue}{rgb}{0.69,0.88,0.90}
\definecolor{RosyBrown1}{rgb}{1.00,0.76,0.76}
\definecolor{RosyBrown2}{rgb}{0.93,0.71,0.71}
\definecolor{RosyBrown3}{rgb}{0.80,0.61,0.61}
\definecolor{RosyBrown4}{rgb}{0.55,0.41,0.41}
\definecolor{RosyBrown}{rgb}{0.74,0.56,0.56}
\definecolor{RoyalBlue1}{rgb}{0.28,0.46,1.00}
\definecolor{RoyalBlue2}{rgb}{0.26,0.43,0.93}
\definecolor{RoyalBlue3}{rgb}{0.23,0.37,0.80}
\definecolor{RoyalBlue4}{rgb}{0.15,0.25,0.55}
\definecolor{RoyalBlue}{rgb}{0.25,0.41,0.88}
\definecolor{SaddleBrown}{rgb}{0.55,0.27,0.07}
\definecolor{SandyBrown}{rgb}{0.96,0.64,0.38}
\definecolor{SeaGreen1}{rgb}{0.33,1.00,0.62}
\definecolor{SeaGreen2}{rgb}{0.31,0.93,0.58}
\definecolor{SeaGreen3}{rgb}{0.26,0.80,0.50}
\definecolor{SeaGreen4}{rgb}{0.18,0.55,0.34}
\definecolor{SeaGreen}{rgb}{0.18,0.55,0.34}
\definecolor{SkyBlue1}{rgb}{0.53,0.81,1.00}
\definecolor{SkyBlue2}{rgb}{0.49,0.75,0.93}
\definecolor{SkyBlue3}{rgb}{0.42,0.65,0.80}
\definecolor{SkyBlue4}{rgb}{0.29,0.44,0.55}
\definecolor{SkyBlue}{rgb}{0.53,0.81,0.92}
\definecolor{SlateBlue1}{rgb}{0.51,0.44,1.00}
\definecolor{SlateBlue2}{rgb}{0.48,0.40,0.93}
\definecolor{SlateBlue3}{rgb}{0.41,0.35,0.80}
\definecolor{SlateBlue4}{rgb}{0.28,0.24,0.55}
\definecolor{SlateBlue}{rgb}{0.42,0.35,0.80}
\definecolor{SlateGray1}{rgb}{0.78,0.89,1.00}
\definecolor{SlateGray2}{rgb}{0.73,0.83,0.93}
\definecolor{SlateGray3}{rgb}{0.62,0.71,0.80}
\definecolor{SlateGray4}{rgb}{0.42,0.48,0.55}
\definecolor{SlateGray}{rgb}{0.44,0.50,0.56}
\definecolor{SlateGrey}{rgb}{0.44,0.50,0.56}
\definecolor{SpringGreen1}{rgb}{0.00,1.00,0.50}
\definecolor{SpringGreen2}{rgb}{0.00,0.93,0.46}
\definecolor{SpringGreen3}{rgb}{0.00,0.80,0.40}
\definecolor{SpringGreen4}{rgb}{0.00,0.55,0.27}
\definecolor{SpringGreen}{rgb}{0.00,1.00,0.50}
\definecolor{SteelBlue1}{rgb}{0.39,0.72,1.00}
\definecolor{SteelBlue2}{rgb}{0.36,0.67,0.93}
\definecolor{SteelBlue3}{rgb}{0.31,0.58,0.80}
\definecolor{SteelBlue4}{rgb}{0.21,0.39,0.55}
\definecolor{SteelBlue}{rgb}{0.27,0.51,0.71}
\definecolor{VioletRed1}{rgb}{1.00,0.24,0.59}
\definecolor{VioletRed2}{rgb}{0.93,0.23,0.55}
\definecolor{VioletRed3}{rgb}{0.80,0.20,0.47}
\definecolor{VioletRed4}{rgb}{0.55,0.13,0.32}
\definecolor{VioletRed}{rgb}{0.82,0.13,0.56}
\definecolor{WhiteSmoke}{rgb}{0.96,0.96,0.96}
\definecolor{YellowGreen}{rgb}{0.60,0.80,0.20}
\definecolor{aliceblue}{rgb}{0.94,0.97,1.00}
\definecolor{antiquewhite}{rgb}{0.98,0.92,0.84}
\definecolor{aquamarine1}{rgb}{0.50,1.00,0.83}
\definecolor{aquamarine2}{rgb}{0.46,0.93,0.78}
\definecolor{aquamarine3}{rgb}{0.40,0.80,0.67}
\definecolor{aquamarine4}{rgb}{0.27,0.55,0.45}
\definecolor{aquamarine}{rgb}{0.50,1.00,0.83}
\definecolor{azure1}{rgb}{0.94,1.00,1.00}
\definecolor{azure2}{rgb}{0.88,0.93,0.93}
\definecolor{azure3}{rgb}{0.76,0.80,0.80}
\definecolor{azure4}{rgb}{0.51,0.55,0.55}
\definecolor{azure}{rgb}{0.94,1.00,1.00}
\definecolor{beige}{rgb}{0.96,0.96,0.86}
\definecolor{bisque1}{rgb}{1.00,0.89,0.77}
\definecolor{bisque2}{rgb}{0.93,0.84,0.72}
\definecolor{bisque3}{rgb}{0.80,0.72,0.62}
\definecolor{bisque4}{rgb}{0.55,0.49,0.42}
\definecolor{bisque}{rgb}{1.00,0.89,0.77}
\definecolor{black}{rgb}{0.00,0.00,0.00}
\definecolor{blanchedalmond}{rgb}{1.00,0.92,0.80}
\definecolor{blue1}{rgb}{0.00,0.00,1.00}
\definecolor{blue2}{rgb}{0.00,0.00,0.93}
\definecolor{blue3}{rgb}{0.00,0.00,0.80}
\definecolor{blue4}{rgb}{0.00,0.00,0.55}
\definecolor{blueviolet}{rgb}{0.54,0.17,0.89}
\definecolor{blue}{rgb}{0.00,0.00,1.00}
\definecolor{brown1}{rgb}{1.00,0.25,0.25}
\definecolor{brown2}{rgb}{0.93,0.23,0.23}
\definecolor{brown3}{rgb}{0.80,0.20,0.20}
\definecolor{brown4}{rgb}{0.55,0.14,0.14}
\definecolor{brown}{rgb}{0.65,0.16,0.16}
\definecolor{burlywood1}{rgb}{1.00,0.83,0.61}
\definecolor{burlywood2}{rgb}{0.93,0.77,0.57}
\definecolor{burlywood3}{rgb}{0.80,0.67,0.49}
\definecolor{burlywood4}{rgb}{0.55,0.45,0.33}
\definecolor{burlywood}{rgb}{0.87,0.72,0.53}
\definecolor{cadetblue}{rgb}{0.37,0.62,0.63}
\definecolor{chartreuse1}{rgb}{0.50,1.00,0.00}
\definecolor{chartreuse2}{rgb}{0.46,0.93,0.00}
\definecolor{chartreuse3}{rgb}{0.40,0.80,0.00}
\definecolor{chartreuse4}{rgb}{0.27,0.55,0.00}
\definecolor{chartreuse}{rgb}{0.50,1.00,0.00}
\definecolor{chocolate1}{rgb}{1.00,0.50,0.14}
\definecolor{chocolate2}{rgb}{0.93,0.46,0.13}
\definecolor{chocolate3}{rgb}{0.80,0.40,0.11}
\definecolor{chocolate4}{rgb}{0.55,0.27,0.07}
\definecolor{chocolate}{rgb}{0.82,0.41,0.12}
\definecolor{coral1}{rgb}{1.00,0.45,0.34}
\definecolor{coral2}{rgb}{0.93,0.42,0.31}
\definecolor{coral3}{rgb}{0.80,0.36,0.27}
\definecolor{coral4}{rgb}{0.55,0.24,0.18}
\definecolor{coral}{rgb}{1.00,0.50,0.31}
\definecolor{cornflowerblue}{rgb}{0.39,0.58,0.93}
\definecolor{cornsilk1}{rgb}{1.00,0.97,0.86}
\definecolor{cornsilk2}{rgb}{0.93,0.91,0.80}
\definecolor{cornsilk3}{rgb}{0.80,0.78,0.69}
\definecolor{cornsilk4}{rgb}{0.55,0.53,0.47}
\definecolor{cornsilk}{rgb}{1.00,0.97,0.86}
\definecolor{cyan1}{rgb}{0.00,1.00,1.00}
\definecolor{cyan2}{rgb}{0.00,0.93,0.93}
\definecolor{cyan3}{rgb}{0.00,0.80,0.80}
\definecolor{cyan4}{rgb}{0.00,0.55,0.55}
\definecolor{cyan}{rgb}{0.00,1.00,1.00}
\definecolor{darkblue}{rgb}{0.00,0.00,0.55}
\definecolor{darkcyan}{rgb}{0.00,0.55,0.55}
\definecolor{darkgoldenrod}{rgb}{0.72,0.53,0.04}
\definecolor{darkgray}{rgb}{0.66,0.66,0.66}
\definecolor{darkgreen}{rgb}{0.00,0.39,0.00}
\definecolor{darkgrey}{rgb}{0.66,0.66,0.66}
\definecolor{darkkhaki}{rgb}{0.74,0.72,0.42}
\definecolor{darkmagenta}{rgb}{0.55,0.00,0.55}
\definecolor{darkolive}{rgb}{0.33,0.42,0.18}
\definecolor{darkorange}{rgb}{1.00,0.55,0.00}
\definecolor{darkorchid}{rgb}{0.60,0.20,0.80}
\definecolor{darkred}{rgb}{0.55,0.00,0.00}
\definecolor{darksalmon}{rgb}{0.91,0.59,0.48}
\definecolor{darksea}{rgb}{0.56,0.74,0.56}
\definecolor{darkslate}{rgb}{0.18,0.31,0.31}
\definecolor{darkslate}{rgb}{0.18,0.31,0.31}
\definecolor{darkslate}{rgb}{0.28,0.24,0.55}
\definecolor{darkturquoise}{rgb}{0.00,0.81,0.82}
\definecolor{darkviolet}{rgb}{0.58,0.00,0.83}
\definecolor{deeppink}{rgb}{1.00,0.08,0.58}
\definecolor{deepsky}{rgb}{0.00,0.75,1.00}
\definecolor{dimgray}{rgb}{0.41,0.41,0.41}
\definecolor{dimgrey}{rgb}{0.41,0.41,0.41}
\definecolor{dodgerblue}{rgb}{0.12,0.56,1.00}
\definecolor{firebrick1}{rgb}{1.00,0.19,0.19}
\definecolor{firebrick2}{rgb}{0.93,0.17,0.17}
\definecolor{firebrick3}{rgb}{0.80,0.15,0.15}
\definecolor{firebrick4}{rgb}{0.55,0.10,0.10}
\definecolor{firebrick}{rgb}{0.70,0.13,0.13}
\definecolor{floralwhite}{rgb}{1.00,0.98,0.94}
\definecolor{forestgreen}{rgb}{0.13,0.55,0.13}
\definecolor{gainsboro}{rgb}{0.86,0.86,0.86}
\definecolor{ghostwhite}{rgb}{0.97,0.97,1.00}
\definecolor{gold1}{rgb}{1.00,0.84,0.00}
\definecolor{gold2}{rgb}{0.93,0.79,0.00}
\definecolor{gold3}{rgb}{0.80,0.68,0.00}
\definecolor{gold4}{rgb}{0.55,0.46,0.00}
\definecolor{goldenrod1}{rgb}{1.00,0.76,0.15}
\definecolor{goldenrod2}{rgb}{0.93,0.71,0.13}
\definecolor{goldenrod3}{rgb}{0.80,0.61,0.11}
\definecolor{goldenrod4}{rgb}{0.55,0.41,0.08}
\definecolor{goldenrod}{rgb}{0.85,0.65,0.13}
\definecolor{gold}{rgb}{1.00,0.84,0.00}
\definecolor{gray0}{rgb}{0.00,0.00,0.00}
\definecolor{gray100}{rgb}{1.00,1.00,1.00}
\definecolor{gray10}{rgb}{0.10,0.10,0.10}
\definecolor{gray11}{rgb}{0.11,0.11,0.11}
\definecolor{gray12}{rgb}{0.12,0.12,0.12}
\definecolor{gray13}{rgb}{0.13,0.13,0.13}
\definecolor{gray14}{rgb}{0.14,0.14,0.14}
\definecolor{gray15}{rgb}{0.15,0.15,0.15}
\definecolor{gray16}{rgb}{0.16,0.16,0.16}
\definecolor{gray17}{rgb}{0.17,0.17,0.17}
\definecolor{gray18}{rgb}{0.18,0.18,0.18}
\definecolor{gray19}{rgb}{0.19,0.19,0.19}
\definecolor{gray1}{rgb}{0.01,0.01,0.01}
\definecolor{gray20}{rgb}{0.20,0.20,0.20}
\definecolor{gray21}{rgb}{0.21,0.21,0.21}
\definecolor{gray22}{rgb}{0.22,0.22,0.22}
\definecolor{gray23}{rgb}{0.23,0.23,0.23}
\definecolor{gray24}{rgb}{0.24,0.24,0.24}
\definecolor{gray25}{rgb}{0.25,0.25,0.25}
\definecolor{gray26}{rgb}{0.26,0.26,0.26}
\definecolor{gray27}{rgb}{0.27,0.27,0.27}
\definecolor{gray28}{rgb}{0.28,0.28,0.28}
\definecolor{gray29}{rgb}{0.29,0.29,0.29}
\definecolor{gray2}{rgb}{0.02,0.02,0.02}
\definecolor{gray30}{rgb}{0.30,0.30,0.30}
\definecolor{gray31}{rgb}{0.31,0.31,0.31}
\definecolor{gray32}{rgb}{0.32,0.32,0.32}
\definecolor{gray33}{rgb}{0.33,0.33,0.33}
\definecolor{gray34}{rgb}{0.34,0.34,0.34}
\definecolor{gray35}{rgb}{0.35,0.35,0.35}
\definecolor{gray36}{rgb}{0.36,0.36,0.36}
\definecolor{gray37}{rgb}{0.37,0.37,0.37}
\definecolor{gray38}{rgb}{0.38,0.38,0.38}
\definecolor{gray39}{rgb}{0.39,0.39,0.39}
\definecolor{gray3}{rgb}{0.03,0.03,0.03}
\definecolor{gray40}{rgb}{0.40,0.40,0.40}
\definecolor{gray41}{rgb}{0.41,0.41,0.41}
\definecolor{gray42}{rgb}{0.42,0.42,0.42}
\definecolor{gray43}{rgb}{0.43,0.43,0.43}
\definecolor{gray44}{rgb}{0.44,0.44,0.44}
\definecolor{gray45}{rgb}{0.45,0.45,0.45}
\definecolor{gray46}{rgb}{0.46,0.46,0.46}
\definecolor{gray47}{rgb}{0.47,0.47,0.47}
\definecolor{gray48}{rgb}{0.48,0.48,0.48}
\definecolor{gray49}{rgb}{0.49,0.49,0.49}
\definecolor{gray4}{rgb}{0.04,0.04,0.04}
\definecolor{gray50}{rgb}{0.50,0.50,0.50}
\definecolor{gray51}{rgb}{0.51,0.51,0.51}
\definecolor{gray52}{rgb}{0.52,0.52,0.52}
\definecolor{gray53}{rgb}{0.53,0.53,0.53}
\definecolor{gray54}{rgb}{0.54,0.54,0.54}
\definecolor{gray55}{rgb}{0.55,0.55,0.55}
\definecolor{gray56}{rgb}{0.56,0.56,0.56}
\definecolor{gray57}{rgb}{0.57,0.57,0.57}
\definecolor{gray58}{rgb}{0.58,0.58,0.58}
\definecolor{gray59}{rgb}{0.59,0.59,0.59}
\definecolor{gray5}{rgb}{0.05,0.05,0.05}
\definecolor{gray60}{rgb}{0.60,0.60,0.60}
\definecolor{gray61}{rgb}{0.61,0.61,0.61}
\definecolor{gray62}{rgb}{0.62,0.62,0.62}
\definecolor{gray63}{rgb}{0.63,0.63,0.63}
\definecolor{gray64}{rgb}{0.64,0.64,0.64}
\definecolor{gray65}{rgb}{0.65,0.65,0.65}
\definecolor{gray66}{rgb}{0.66,0.66,0.66}
\definecolor{gray67}{rgb}{0.67,0.67,0.67}
\definecolor{gray68}{rgb}{0.68,0.68,0.68}
\definecolor{gray69}{rgb}{0.69,0.69,0.69}
\definecolor{gray6}{rgb}{0.06,0.06,0.06}
\definecolor{gray70}{rgb}{0.70,0.70,0.70}
\definecolor{gray71}{rgb}{0.71,0.71,0.71}
\definecolor{gray72}{rgb}{0.72,0.72,0.72}
\definecolor{gray73}{rgb}{0.73,0.73,0.73}
\definecolor{gray74}{rgb}{0.74,0.74,0.74}
\definecolor{gray75}{rgb}{0.75,0.75,0.75}
\definecolor{gray76}{rgb}{0.76,0.76,0.76}
\definecolor{gray77}{rgb}{0.77,0.77,0.77}
\definecolor{gray78}{rgb}{0.78,0.78,0.78}
\definecolor{gray79}{rgb}{0.79,0.79,0.79}
\definecolor{gray7}{rgb}{0.07,0.07,0.07}
\definecolor{gray80}{rgb}{0.80,0.80,0.80}
\definecolor{gray81}{rgb}{0.81,0.81,0.81}
\definecolor{gray82}{rgb}{0.82,0.82,0.82}
\definecolor{gray83}{rgb}{0.83,0.83,0.83}
\definecolor{gray84}{rgb}{0.84,0.84,0.84}
\definecolor{gray85}{rgb}{0.85,0.85,0.85}
\definecolor{gray86}{rgb}{0.86,0.86,0.86}
\definecolor{gray87}{rgb}{0.87,0.87,0.87}
\definecolor{gray88}{rgb}{0.88,0.88,0.88}
\definecolor{gray89}{rgb}{0.89,0.89,0.89}
\definecolor{gray8}{rgb}{0.08,0.08,0.08}
\definecolor{gray90}{rgb}{0.90,0.90,0.90}
\definecolor{gray91}{rgb}{0.91,0.91,0.91}
\definecolor{gray92}{rgb}{0.92,0.92,0.92}
\definecolor{gray93}{rgb}{0.93,0.93,0.93}
\definecolor{gray94}{rgb}{0.94,0.94,0.94}
\definecolor{gray95}{rgb}{0.95,0.95,0.95}
\definecolor{gray96}{rgb}{0.96,0.96,0.96}
\definecolor{gray97}{rgb}{0.97,0.97,0.97}
\definecolor{gray98}{rgb}{0.98,0.98,0.98}
\definecolor{gray99}{rgb}{0.99,0.99,0.99}
\definecolor{gray9}{rgb}{0.09,0.09,0.09}
\definecolor{gray}{rgb}{0.75,0.75,0.75}
\definecolor{green1}{rgb}{0.00,1.00,0.00}
\definecolor{green2}{rgb}{0.00,0.93,0.00}
\definecolor{green3}{rgb}{0.00,0.80,0.00}
\definecolor{green4}{rgb}{0.00,0.55,0.00}
\definecolor{greenyellow}{rgb}{0.68,1.00,0.18}
\definecolor{green}{rgb}{0.00,1.00,0.00}
\definecolor{grey0}{rgb}{0.00,0.00,0.00}
\definecolor{grey100}{rgb}{1.00,1.00,1.00}
\definecolor{grey10}{rgb}{0.10,0.10,0.10}
\definecolor{grey11}{rgb}{0.11,0.11,0.11}
\definecolor{grey12}{rgb}{0.12,0.12,0.12}
\definecolor{grey13}{rgb}{0.13,0.13,0.13}
\definecolor{grey14}{rgb}{0.14,0.14,0.14}
\definecolor{grey15}{rgb}{0.15,0.15,0.15}
\definecolor{grey16}{rgb}{0.16,0.16,0.16}
\definecolor{grey17}{rgb}{0.17,0.17,0.17}
\definecolor{grey18}{rgb}{0.18,0.18,0.18}
\definecolor{grey19}{rgb}{0.19,0.19,0.19}
\definecolor{grey1}{rgb}{0.01,0.01,0.01}
\definecolor{grey20}{rgb}{0.20,0.20,0.20}
\definecolor{grey21}{rgb}{0.21,0.21,0.21}
\definecolor{grey22}{rgb}{0.22,0.22,0.22}
\definecolor{grey23}{rgb}{0.23,0.23,0.23}
\definecolor{grey24}{rgb}{0.24,0.24,0.24}
\definecolor{grey25}{rgb}{0.25,0.25,0.25}
\definecolor{grey26}{rgb}{0.26,0.26,0.26}
\definecolor{grey27}{rgb}{0.27,0.27,0.27}
\definecolor{grey28}{rgb}{0.28,0.28,0.28}
\definecolor{grey29}{rgb}{0.29,0.29,0.29}
\definecolor{grey2}{rgb}{0.02,0.02,0.02}
\definecolor{grey30}{rgb}{0.30,0.30,0.30}
\definecolor{grey31}{rgb}{0.31,0.31,0.31}
\definecolor{grey32}{rgb}{0.32,0.32,0.32}
\definecolor{grey33}{rgb}{0.33,0.33,0.33}
\definecolor{grey34}{rgb}{0.34,0.34,0.34}
\definecolor{grey35}{rgb}{0.35,0.35,0.35}
\definecolor{grey36}{rgb}{0.36,0.36,0.36}
\definecolor{grey37}{rgb}{0.37,0.37,0.37}
\definecolor{grey38}{rgb}{0.38,0.38,0.38}
\definecolor{grey39}{rgb}{0.39,0.39,0.39}
\definecolor{grey3}{rgb}{0.03,0.03,0.03}
\definecolor{grey40}{rgb}{0.40,0.40,0.40}
\definecolor{grey41}{rgb}{0.41,0.41,0.41}
\definecolor{grey42}{rgb}{0.42,0.42,0.42}
\definecolor{grey43}{rgb}{0.43,0.43,0.43}
\definecolor{grey44}{rgb}{0.44,0.44,0.44}
\definecolor{grey45}{rgb}{0.45,0.45,0.45}
\definecolor{grey46}{rgb}{0.46,0.46,0.46}
\definecolor{grey47}{rgb}{0.47,0.47,0.47}
\definecolor{grey48}{rgb}{0.48,0.48,0.48}
\definecolor{grey49}{rgb}{0.49,0.49,0.49}
\definecolor{grey4}{rgb}{0.04,0.04,0.04}
\definecolor{grey50}{rgb}{0.50,0.50,0.50}
\definecolor{grey51}{rgb}{0.51,0.51,0.51}
\definecolor{grey52}{rgb}{0.52,0.52,0.52}
\definecolor{grey53}{rgb}{0.53,0.53,0.53}
\definecolor{grey54}{rgb}{0.54,0.54,0.54}
\definecolor{grey55}{rgb}{0.55,0.55,0.55}
\definecolor{grey56}{rgb}{0.56,0.56,0.56}
\definecolor{grey57}{rgb}{0.57,0.57,0.57}
\definecolor{grey58}{rgb}{0.58,0.58,0.58}
\definecolor{grey59}{rgb}{0.59,0.59,0.59}
\definecolor{grey5}{rgb}{0.05,0.05,0.05}
\definecolor{grey60}{rgb}{0.60,0.60,0.60}
\definecolor{grey61}{rgb}{0.61,0.61,0.61}
\definecolor{grey62}{rgb}{0.62,0.62,0.62}
\definecolor{grey63}{rgb}{0.63,0.63,0.63}
\definecolor{grey64}{rgb}{0.64,0.64,0.64}
\definecolor{grey65}{rgb}{0.65,0.65,0.65}
\definecolor{grey66}{rgb}{0.66,0.66,0.66}
\definecolor{grey67}{rgb}{0.67,0.67,0.67}
\definecolor{grey68}{rgb}{0.68,0.68,0.68}
\definecolor{grey69}{rgb}{0.69,0.69,0.69}
\definecolor{grey6}{rgb}{0.06,0.06,0.06}
\definecolor{grey70}{rgb}{0.70,0.70,0.70}
\definecolor{grey71}{rgb}{0.71,0.71,0.71}
\definecolor{grey72}{rgb}{0.72,0.72,0.72}
\definecolor{grey73}{rgb}{0.73,0.73,0.73}
\definecolor{grey74}{rgb}{0.74,0.74,0.74}
\definecolor{grey75}{rgb}{0.75,0.75,0.75}
\definecolor{grey76}{rgb}{0.76,0.76,0.76}
\definecolor{grey77}{rgb}{0.77,0.77,0.77}
\definecolor{grey78}{rgb}{0.78,0.78,0.78}
\definecolor{grey79}{rgb}{0.79,0.79,0.79}
\definecolor{grey7}{rgb}{0.07,0.07,0.07}
\definecolor{grey80}{rgb}{0.80,0.80,0.80}
\definecolor{grey81}{rgb}{0.81,0.81,0.81}
\definecolor{grey82}{rgb}{0.82,0.82,0.82}
\definecolor{grey83}{rgb}{0.83,0.83,0.83}
\definecolor{grey84}{rgb}{0.84,0.84,0.84}
\definecolor{grey85}{rgb}{0.85,0.85,0.85}
\definecolor{grey86}{rgb}{0.86,0.86,0.86}
\definecolor{grey87}{rgb}{0.87,0.87,0.87}
\definecolor{grey88}{rgb}{0.88,0.88,0.88}
\definecolor{grey89}{rgb}{0.89,0.89,0.89}
\definecolor{grey8}{rgb}{0.08,0.08,0.08}
\definecolor{grey90}{rgb}{0.90,0.90,0.90}
\definecolor{grey91}{rgb}{0.91,0.91,0.91}
\definecolor{grey92}{rgb}{0.92,0.92,0.92}
\definecolor{grey93}{rgb}{0.93,0.93,0.93}
\definecolor{grey94}{rgb}{0.94,0.94,0.94}
\definecolor{grey95}{rgb}{0.95,0.95,0.95}
\definecolor{grey96}{rgb}{0.96,0.96,0.96}
\definecolor{grey97}{rgb}{0.97,0.97,0.97}
\definecolor{grey98}{rgb}{0.98,0.98,0.98}
\definecolor{grey99}{rgb}{0.99,0.99,0.99}
\definecolor{grey9}{rgb}{0.09,0.09,0.09}
\definecolor{grey}{rgb}{0.75,0.75,0.75}
\definecolor{honeydew1}{rgb}{0.94,1.00,0.94}
\definecolor{honeydew2}{rgb}{0.88,0.93,0.88}
\definecolor{honeydew3}{rgb}{0.76,0.80,0.76}
\definecolor{honeydew4}{rgb}{0.51,0.55,0.51}
\definecolor{honeydew}{rgb}{0.94,1.00,0.94}
\definecolor{hotpink}{rgb}{1.00,0.41,0.71}
\definecolor{indianred}{rgb}{0.80,0.36,0.36}
\definecolor{ivory1}{rgb}{1.00,1.00,0.94}
\definecolor{ivory2}{rgb}{0.93,0.93,0.88}
\definecolor{ivory3}{rgb}{0.80,0.80,0.76}
\definecolor{ivory4}{rgb}{0.55,0.55,0.51}
\definecolor{ivory}{rgb}{1.00,1.00,0.94}
\definecolor{khaki1}{rgb}{1.00,0.96,0.56}
\definecolor{khaki2}{rgb}{0.93,0.90,0.52}
\definecolor{khaki3}{rgb}{0.80,0.78,0.45}
\definecolor{khaki4}{rgb}{0.55,0.53,0.31}
\definecolor{khaki}{rgb}{0.94,0.90,0.55}
\definecolor{lavenderblush}{rgb}{1.00,0.94,0.96}
\definecolor{lavender}{rgb}{0.90,0.90,0.98}
\definecolor{lawngreen}{rgb}{0.49,0.99,0.00}
\definecolor{lemonchiffon}{rgb}{1.00,0.98,0.80}
\definecolor{lightblue}{rgb}{0.68,0.85,0.90}
\definecolor{lightcoral}{rgb}{0.94,0.50,0.50}
\definecolor{lightcyan}{rgb}{0.88,1.00,1.00}
\definecolor{lightgoldenrod}{rgb}{0.93,0.87,0.51}
\definecolor{lightgoldenrod}{rgb}{0.98,0.98,0.82}
\definecolor{lightgray}{rgb}{0.83,0.83,0.83}
\definecolor{lightgreen}{rgb}{0.56,0.93,0.56}
\definecolor{lightgrey}{rgb}{0.83,0.83,0.83}
\definecolor{lightpink}{rgb}{1.00,0.71,0.76}
\definecolor{lightsalmon}{rgb}{1.00,0.63,0.48}
\definecolor{lightsea}{rgb}{0.13,0.70,0.67}
\definecolor{lightsky}{rgb}{0.53,0.81,0.98}
\definecolor{lightslate}{rgb}{0.47,0.53,0.60}
\definecolor{lightslate}{rgb}{0.47,0.53,0.60}
\definecolor{lightslate}{rgb}{0.52,0.44,1.00}
\definecolor{lightsteel}{rgb}{0.69,0.77,0.87}
\definecolor{lightyellow}{rgb}{1.00,1.00,0.88}
\definecolor{limegreen}{rgb}{0.20,0.80,0.20}
\definecolor{linen}{rgb}{0.98,0.94,0.90}
\definecolor{magenta1}{rgb}{1.00,0.00,1.00}
\definecolor{magenta2}{rgb}{0.93,0.00,0.93}
\definecolor{magenta3}{rgb}{0.80,0.00,0.80}
\definecolor{magenta4}{rgb}{0.55,0.00,0.55}
\definecolor{magenta}{rgb}{1.00,0.00,1.00}
\definecolor{maroon1}{rgb}{1.00,0.20,0.70}
\definecolor{maroon2}{rgb}{0.93,0.19,0.65}
\definecolor{maroon3}{rgb}{0.80,0.16,0.56}
\definecolor{maroon4}{rgb}{0.55,0.11,0.38}
\definecolor{maroon}{rgb}{0.69,0.19,0.38}
\definecolor{mediumaquamarine}{rgb}{0.40,0.80,0.67}
\definecolor{mediumblue}{rgb}{0.00,0.00,0.80}
\definecolor{mediumorchid}{rgb}{0.73,0.33,0.83}
\definecolor{mediumpurple}{rgb}{0.58,0.44,0.86}
\definecolor{mediumsea}{rgb}{0.24,0.70,0.44}
\definecolor{mediumslate}{rgb}{0.48,0.41,0.93}
\definecolor{mediumspring}{rgb}{0.00,0.98,0.60}
\definecolor{mediumturquoise}{rgb}{0.28,0.82,0.80}
\definecolor{mediumviolet}{rgb}{0.78,0.08,0.52}
\definecolor{midnightblue}{rgb}{0.10,0.10,0.44}
\definecolor{mintcream}{rgb}{0.96,1.00,0.98}
\definecolor{mistyrose}{rgb}{1.00,0.89,0.88}
\definecolor{moccasin}{rgb}{1.00,0.89,0.71}
\definecolor{navajowhite}{rgb}{1.00,0.87,0.68}
\definecolor{navyblue}{rgb}{0.00,0.00,0.50}
\definecolor{navy}{rgb}{0.00,0.00,0.50}
\definecolor{oldlace}{rgb}{0.99,0.96,0.90}
\definecolor{olivedrab}{rgb}{0.42,0.56,0.14}
\definecolor{orange1}{rgb}{1.00,0.65,0.00}
\definecolor{orange2}{rgb}{0.93,0.60,0.00}
\definecolor{orange3}{rgb}{0.80,0.52,0.00}
\definecolor{orange4}{rgb}{0.55,0.35,0.00}
\definecolor{orangered}{rgb}{1.00,0.27,0.00}
\definecolor{orange}{rgb}{1.00,0.65,0.00}
\definecolor{orchid1}{rgb}{1.00,0.51,0.98}
\definecolor{orchid2}{rgb}{0.93,0.48,0.91}
\definecolor{orchid3}{rgb}{0.80,0.41,0.79}
\definecolor{orchid4}{rgb}{0.55,0.28,0.54}
\definecolor{orchid}{rgb}{0.85,0.44,0.84}
\definecolor{palegoldenrod}{rgb}{0.93,0.91,0.67}
\definecolor{palegreen}{rgb}{0.60,0.98,0.60}
\definecolor{paleturquoise}{rgb}{0.69,0.93,0.93}
\definecolor{paleviolet}{rgb}{0.86,0.44,0.58}
\definecolor{papayawhip}{rgb}{1.00,0.94,0.84}
\definecolor{peachpuff}{rgb}{1.00,0.85,0.73}
\definecolor{peru}{rgb}{0.80,0.52,0.25}
\definecolor{pink1}{rgb}{1.00,0.71,0.77}
\definecolor{pink2}{rgb}{0.93,0.66,0.72}
\definecolor{pink3}{rgb}{0.80,0.57,0.62}
\definecolor{pink4}{rgb}{0.55,0.39,0.42}
\definecolor{pink}{rgb}{1.00,0.75,0.80}
\definecolor{plum1}{rgb}{1.00,0.73,1.00}
\definecolor{plum2}{rgb}{0.93,0.68,0.93}
\definecolor{plum3}{rgb}{0.80,0.59,0.80}
\definecolor{plum4}{rgb}{0.55,0.40,0.55}
\definecolor{plum}{rgb}{0.87,0.63,0.87}
\definecolor{powderblue}{rgb}{0.69,0.88,0.90}
\definecolor{purple1}{rgb}{0.61,0.19,1.00}
\definecolor{purple2}{rgb}{0.57,0.17,0.93}
\definecolor{purple3}{rgb}{0.49,0.15,0.80}
\definecolor{purple4}{rgb}{0.33,0.10,0.55}
\definecolor{purple}{rgb}{0.63,0.13,0.94}
\definecolor{red1}{rgb}{1.00,0.00,0.00}
\definecolor{red2}{rgb}{0.93,0.00,0.00}
\definecolor{red3}{rgb}{0.80,0.00,0.00}
\definecolor{red4}{rgb}{0.55,0.00,0.00}
\definecolor{red}{rgb}{1.00,0.00,0.00}
\definecolor{rosybrown}{rgb}{0.74,0.56,0.56}
\definecolor{royalblue}{rgb}{0.25,0.41,0.88}
\definecolor{saddlebrown}{rgb}{0.55,0.27,0.07}
\definecolor{salmon1}{rgb}{1.00,0.55,0.41}
\definecolor{salmon2}{rgb}{0.93,0.51,0.38}
\definecolor{salmon3}{rgb}{0.80,0.44,0.33}
\definecolor{salmon4}{rgb}{0.55,0.30,0.22}
\definecolor{salmon}{rgb}{0.98,0.50,0.45}
\definecolor{sandybrown}{rgb}{0.96,0.64,0.38}
\definecolor{seagreen}{rgb}{0.18,0.55,0.34}
\definecolor{seashell1}{rgb}{1.00,0.96,0.93}
\definecolor{seashell2}{rgb}{0.93,0.90,0.87}
\definecolor{seashell3}{rgb}{0.80,0.77,0.75}
\definecolor{seashell4}{rgb}{0.55,0.53,0.51}
\definecolor{seashell}{rgb}{1.00,0.96,0.93}
\definecolor{sienna1}{rgb}{1.00,0.51,0.28}
\definecolor{sienna2}{rgb}{0.93,0.47,0.26}
\definecolor{sienna3}{rgb}{0.80,0.41,0.22}
\definecolor{sienna4}{rgb}{0.55,0.28,0.15}
\definecolor{sienna}{rgb}{0.63,0.32,0.18}
\definecolor{skyblue}{rgb}{0.53,0.81,0.92}
\definecolor{slateblue}{rgb}{0.42,0.35,0.80}
\definecolor{slategray}{rgb}{0.44,0.50,0.56}
\definecolor{slategrey}{rgb}{0.44,0.50,0.56}
\definecolor{snow1}{rgb}{1.00,0.98,0.98}
\definecolor{snow2}{rgb}{0.93,0.91,0.91}
\definecolor{snow3}{rgb}{0.80,0.79,0.79}
\definecolor{snow4}{rgb}{0.55,0.54,0.54}
\definecolor{snow}{rgb}{1.00,0.98,0.98}
\definecolor{springgreen}{rgb}{0.00,1.00,0.50}
\definecolor{steelblue}{rgb}{0.27,0.51,0.71}
\definecolor{tan1}{rgb}{1.00,0.65,0.31}
\definecolor{tan2}{rgb}{0.93,0.60,0.29}
\definecolor{tan3}{rgb}{0.80,0.52,0.25}
\definecolor{tan4}{rgb}{0.55,0.35,0.17}
\definecolor{tan}{rgb}{0.82,0.71,0.55}
\definecolor{thistle1}{rgb}{1.00,0.88,1.00}
\definecolor{thistle2}{rgb}{0.93,0.82,0.93}
\definecolor{thistle3}{rgb}{0.80,0.71,0.80}
\definecolor{thistle4}{rgb}{0.55,0.48,0.55}
\definecolor{thistle}{rgb}{0.85,0.75,0.85}
\definecolor{tomato1}{rgb}{1.00,0.39,0.28}
\definecolor{tomato2}{rgb}{0.93,0.36,0.26}
\definecolor{tomato3}{rgb}{0.80,0.31,0.22}
\definecolor{tomato4}{rgb}{0.55,0.21,0.15}
\definecolor{tomato}{rgb}{1.00,0.39,0.28}
\definecolor{turquoise1}{rgb}{0.00,0.96,1.00}
\definecolor{turquoise2}{rgb}{0.00,0.90,0.93}
\definecolor{turquoise3}{rgb}{0.00,0.77,0.80}
\definecolor{turquoise4}{rgb}{0.00,0.53,0.55}
\definecolor{turquoise}{rgb}{0.25,0.88,0.82}
\definecolor{violetred}{rgb}{0.82,0.13,0.56}
\definecolor{violet}{rgb}{0.93,0.51,0.93}
\definecolor{wheat1}{rgb}{1.00,0.91,0.73}
\definecolor{wheat2}{rgb}{0.93,0.85,0.68}
\definecolor{wheat3}{rgb}{0.80,0.73,0.59}
\definecolor{wheat4}{rgb}{0.55,0.49,0.40}
\definecolor{wheat}{rgb}{0.96,0.87,0.70}
\definecolor{whitesmoke}{rgb}{0.96,0.96,0.96}
\definecolor{white}{rgb}{1.00,1.00,1.00}
\definecolor{yellow1}{rgb}{1.00,1.00,0.00}
\definecolor{yellow2}{rgb}{0.93,0.93,0.00}
\definecolor{yellow3}{rgb}{0.80,0.80,0.00}
\definecolor{yellow4}{rgb}{0.55,0.55,0.00}
\definecolor{yellowgreen}{rgb}{0.60,0.80,0.20}
\definecolor{yellow}{rgb}{1.00,1.00,0.00}

\usepackage{slashed}

\declareslashed{}{/}{0}{-0.05}{E}
\declareslashed{}{/}{0}{-0.05}{P}
\declareslashed{}{/}{0}{-0.10}{\Widehat{P}}

\def\fsu5{$\cal{F}$-$SU(5)$}
\def\bfsu5{$\boldsymbol{\mathcal{F}}$-$\boldsymbol{SU(5)}$}

\def\m1half{$M_{1/2}$}
\def\m3half{$M_{3/2}$}
\def\m32{$M_{32}$}

\def\mt2{$M_{T2}$}
\def\x2{$\chi^2$}
\def\2b{$M_{T2}b$}

\def\bs0{$B_S^0 \rightarrow \mu^+ \mu^-$}
\newcommand{\ttbar}{t \bar{t}}
\def\met{{\slashed{E}} {}_{\rm T}}

\begin{document}

\title{The 14 TeV LHC Takes Aim at SUSY: \\ A No-Scale Supergravity Model for LHC Run 2 \footnote{Invited Comment for {\it Gravity, Supergravity, and Fundamental Physics: The Richard Arnowitt Symposium}, published in Physica Scripta by the Royal Swedish Academy of Sciences.}}

\author{Tianjun Li}

\affiliation{State Key Laboratory of Theoretical Physics and Kavli Institute for Theoretical Physics China (KITPC),
Institute of Theoretical Physics, Chinese Academy of Sciences, Beijing 100190, P. R. China}

\affiliation{School of Physical Electronics, University of Electronic Science and Technology of China, 
Chengdu 610054, P. R. China }

\author{James A. Maxin}

\affiliation{Department of Physics and Engineering Physics, The University of Tulsa, Tulsa, OK 74104 USA}

\author{Dimitri V. Nanopoulos}

\affiliation{George P. and Cynthia W. Mitchell Institute for Fundamental Physics and Astronomy, Texas A$\&$M University, College Station, TX 77843, USA}

\affiliation{Astroparticle Physics Group, Houston Advanced Research Center (HARC), Mitchell Campus, Woodlands, TX 77381, USA}

\affiliation{Academy of Athens, Division of Natural Sciences, 28 Panepistimiou Avenue, Athens 10679, Greece}

\author{Joel W. Walker}

\affiliation{Department of Physics, Sam Houston State University, Huntsville, TX 77341, USA}


\begin{abstract}

The Supergravity model named No-Scale \fsu5, which is based upon the flipped $SU$(5) Grand Unified Theory (GUT) with additional TeV-scale vector-like flippon multiplets, has been partially probed during the LHC Run 1 at 7--8 TeV, though the majority of its model space remains viable and should be accessible by the 13--14 TeV LHC during Run 2. The model framework possesses the rather unique capacity to provide a light CP-even Higgs boson mass in the favored 124--126 GeV window while simultaneously retaining a testably light supersymmetry (SUSY) spectrum. We summarize the outlook for No-Scale \fsu5 at the 13--14 TeV LHC and review a promising methodology for the discrimination of its long-chain cascade decay signature. We further show that proportional dependence of all model scales upon the unified gaugino mass $M_{1/2}$ minimizes electroweak fine-tuning, allowing the $Z$-boson mass $M_Z$ to be expressed as an explicit function of $M_{1/2}$, $M_Z^2 = M_Z^2 (M_{1/2}^2)$, with implicit dependence upon a dimensionless ratio $c$ of the supersymmetric Higgs mixing parameter $\mu$ and $M_{1/2}$. Finally, we elucidate an empirical connection between recent scalar tensor measurements and No-Scale Supergravity cosmological models that mimic the Starobinsky model of inflation.

\end{abstract}


\pacs{11.10.Kk, 11.25.Mj, 11.25.-w, 12.60.Jv}

\preprint{ACT-05-15}

\maketitle


{\it In memory of Richard Arnowitt, a true giant.  And of Tristan Leggett, a friend and colleague taken too soon.}

\section{Introduction}

The Large Hadron Collider (LHC) Run 2 shall commence soon, fresh off the historic success of the 125~GeV Higgs Boson
discovery~\cite{:2012gk,:2012gu} during the 7--8~TeV Run 1. The primary target for the impending 13--14~TeV phase is
supersymmetry (SUSY), an elegant approach to stabilizing the electroweak scale,
which also provides a favorable cold dark matter candidate,
a mechanism for radiative electroweak symmetry breaking (EWSB),
and a framework whose local extension naturally points the way toward a quantum theory of gravity.
Though no significant signal directly associable with SUSY particles (sparticles) emerged from the 7--8 TeV proton--proton collision,
the escalation to a 13--14 TeV center-of-mass energy will dramatically extend the
mass reach for SUSY pair production events, and enhance the probability of reconstructing any subsequent decay cascade chain.

The absence of a definitive light SUSY signal during LHC Run 1 has been a source of some disappointment amongst SUSY enthusiasts,
although it is axiomatic that failure generally precedes success, and that many dead-ends must often be eliminated before the path is found.
Indeed, the landscape of prospective SUSY models has been so generously populated with attractive candidates,
that substantial winnowing of the chaff is an inevitable component of any efforts directed toward discovery.
Given that the original collection of proposed models has been diminished considerably,
the remaining viable models now assume an elevated significance.

The intent of this work is to discuss one such favored SUSY GUT framework, referred to as No-Scale \fsu5~\cite{
Li:2010ws, Li:2010mi,Li:2010uu,Maxin:2011hy, Li:2011xu,Li:2011ab,Li:2012jf,Li:2013hpa,Li:2013naa,Li:2013bxh,Li:2013mwa,Leggett:2014mza,Leggett:2014hha}.
No-Scale \fsu5 is built upon a tripodal foundation of
the dynamically established No-Scale Supergravity boundary conditions,
the Flipped $SU(5)$ Grand Unified Theory (GUT),
and a pair of TeV-scale hypothetical ``{\it flippon}'' vector-like super-multiplets
derived within local F-theory model building. Remarkably, the confluence of these three concepts resolves several
longstanding theoretical dilemmas, while comparing well with real world experimental observations.

The minimalistic formalism of No-Scale Supergravity~\cite{Cremmer:1983bf,Ellis:1983sf, Ellis:1983ei, Ellis:1984bm, Lahanas:1986uc}
provides for a fundamental connection to string theory in the infrared limit, the natural incorporation of general coordinate
invariance (general relativity), a mechanism for SUSY breaking that preserves a vanishing cosmological constant at the tree level
(facilitating the observed longevity and cosmological flatness of our Universe~\cite{Cremmer:1983bf}), natural suppression of CP
violation and flavor-changing neutral currents, dynamic stabilization of the compactified spacetime by minimization of the
loop-corrected scalar potential, and an extremely economical reduction in parameterization freedom. The split-unification
structure of flipped $SU(5)$~\cite{Nanopoulos:2002qk,Barr:1981qv,Derendinger:1983aj,Antoniadis:1987dx} provides for
fundamental GUT scale Higgs representations (not adjoints), natural doublet-triplet splitting, suppression of dimension-five
proton decay~\cite{Antoniadis:1987dx,Harnik:2004yp}, and a two-step see-saw mechanism for neutrino masses~\cite{Ellis:1992nq,Ellis:1993ks}.
Revisions to the one-loop gauge $\beta$-function coefficients $b_i$ induced by inclusion of the vector-like flippon multiplets create an
essential flattening of the $SU(3)$ renormalization group equation (RGE) running ($b_3 = 0$)~\cite{Li:2010ws}, which translates into an
expanded separation between the primary $SU(3)_C \times SU(2)_L$ unification near $10^{16}$~GeV and the secondary $SU(5) \times U(1)_X$
unification near the Planck mass.  The corresponding baseline extension for logarithmic running of the No-Scale boundary conditions,
especially that imposed ($B_\mu = 0$) on the soft SUSY breaking term $B_\mu$ associated with the Higgs bilinear mass mixing $\mu$,
allows ample space for natural dynamic evolution into phenomenologically favorable values at the electroweak scale. Correlated
flattening of the color-charged gaugino mass scale generates a distinctive SUSY mass pattern of $M(\widetilde{t}_1) < M(\widetilde{g}) < M(\widetilde{q})$,
characterized by a light stop and gluino that are lighter than all other squarks~\cite{Li:2011ab}.

The {\it big} gauge hierarchy problem is resolved by SUSY via logarithmically sequestering the reference to ultra-heavy
(Grand Unification, Planck, String) scales of new physics. Nonetheless, a residual {\it little} hierarchy problem persists,
implicit in the gap separating TeV-scale collider bounds on (strong production of) yet elusive colored sparticle fields.
The massive SUSY scale also seems required for the necessary loop contributions to the physical Higgs mass itself.
One potential mechanism for reconciliation of these considerations without an unnatural appeal to fine tuning, {\it vis-\`a-vis}
unmotivated cancellation of more than (say) a few parts {\it per centum} between contributions to physics at the electroweak
(EW) scale, could be the parsimony of a unified framework wherein the entire physical spectrum (Standard Model + SUSY) may be
expressed as functions of a single parameter.  No-Scale \fsu5 has demonstrated phenomenological
evidence~\cite{Leggett:2014mza,Leggett:2014hha} of such a suppression in the demand for electroweak fine-tuning.

A recent analysis~\cite{Ellis:2013xoa,Ellis:2013nxa,Ellis:2013nka} suggests that a cosmological model based upon the No-Scale
supergravity sector yields compatibility with the Planck satellite measurements. With convenient superpotential parameter choices,
the new cosmological model compatible with Planck data is a No-Scale supergravity realization of the
Starobinsky model of inflation~\cite{Starobinsky:1980te,Mukhanov:1981xt,Starobinsky:1983zz}. We shall elaborate here upon this
intriguing connection between the No-Scale \fsu5 GUT model and a No-Scale Wess-Zumino model of inflation.

\section{The No-Scale \bfsu5 Model}

Supersymmetry naturally solves
the gauge hierarchy problem in the SM, and suggests (given $R$ parity conservation)
the LSP as a suitable cold dark matter candidate.
However, since we do not see mass degeneracy of the superpartners,
SUSY must be broken around the TeV scale. In GUTs with
gravity mediated supersymmetry breaking, called
the supergravity models,
we can fully characterize the supersymmetry breaking
soft terms by four universal parameters
(gaugino mass $M_{1/2}$, scalar mass $M_0$, trilinear soft term $A$, and
the low energy ratio of Higgs vacuum expectation values (VEVs) $\tan\beta$),
plus the sign of the Higgs bilinear mass term $\mu$.

No-Scale Supergravity was proposed~\cite{Cremmer:1983bf}
to address the cosmological flatness problem,
as the subset of supergravity models
which satisfy the following three constraints:
i) the vacuum energy vanishes automatically due to the suitable
 K\"ahler potential; ii) at the minimum of the scalar
potential there exist flat directions that leave the
gravitino mass $M_{3/2}$ undetermined; iii) the quantity
${\rm Str} {\cal M}^2$ is zero at the minimum. If the third condition
were not true, large one-loop corrections would force $M_{3/2}$ to be
either identically zero or of the Planck scale. A simple K\"ahler potential that 
satisfies the first two conditions is~\cite{Ellis:1984bm,Cremmer:1983bf}
\begin{eqnarray} 
K &=& -3 {\rm ln}( T+\overline{T}-\sum_i \overline{\Phi}_i
\Phi_i)~,~
\label{NS-Kahler}
\end{eqnarray}
where $T$ is a modulus field and $\Phi_i$ are matter fields, which parameterize the non-compact $SU(N,1)/SU(N) \times U(1)$ coset space.
The third condition is model dependent and can always be satisfied in
principle~\cite{Ferrara:1994kg}.
For the simple K\"ahler potential in Eq.~(\ref{NS-Kahler})
we automatically obtain the No-Scale boundary condition
$M_0=A=B_{\mu}=0$ while $M_{1/2}$ is allowed,
and indeed required for SUSY breaking.
Because the minimum of the electroweak (EW) Higgs potential
$(V_{EW})_{min}$ depends on $M_{3/2}$,  the gravitino mass is 
determined by the equation $d(V_{EW})_{min}/dM_{3/2}=0$.
Thus, the supersymmetry breaking scale is determined
dynamically. No-Scale supergravity can be
realized in the compactification of the weakly coupled
heterotic string theory~\cite{Witten:1985xb} and the compactification of
M-theory on $S^1/Z_2$ at the leading order~\cite{Li:1997sk}.

In order to achieve true string-scale gauge coupling unification
while avoiding the Landau pole problem,
we supplement the standard ${\cal F}$-lipped $SU(5)\times U(1)_X$~\cite{Nanopoulos:2002qk,Barr:1981qv,Derendinger:1983aj,Antoniadis:1987dx}
SUSY field content with the following TeV-scale vector-like multiplets (flippons)~\cite{Jiang:2006hf}
\begin{eqnarray}
\hspace{-.3in}
& \left( {XF}_{\mathbf{(10,1)}} \equiv (XQ,XD^c,XN^c),~{\overline{XF}}_{\mathbf{({\overline{10}},-1)}} \right)\, ,&
\nonumber \\
\hspace{-.3in}
& \left( {Xl}_{\mathbf{(1, -5)}},~{\overline{Xl}}_{\mathbf{(1, 5)}}\equiv XE^c \right)\, ,&
\label{z1z2}
\end{eqnarray}
where $XQ$, $XD^c$, $XE^c$, $XN^c$ have the same quantum numbers as the
quark doublet, the right-handed down-type quark, charged lepton, and
neutrino, respectively.
Such kind of models can be realized in ${\cal F}$-ree ${\cal F}$-ermionic string
constructions~\cite{Lopez:1992kg},
and ${\cal F}$-theory model building~\cite{Jiang:2009zza,Jiang:2009za}. Thus, they have been 
dubbed ${\cal F}$-$SU(5)$~\cite{Jiang:2009zza}.

\section{The Wedge of Bare-Minimal Constraints}

\begin{figure}[htp]
        \centering
        \includegraphics[width=0.5\textwidth]{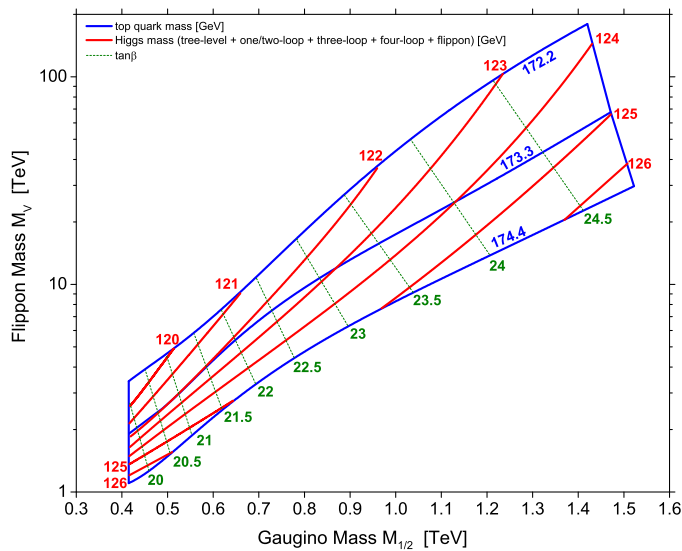}
        \caption{Constrained model space of No-Scale \fsu5 as a function of the gaugino mass $M_{1/2}$ and flippon mass $M_V$. The thick lines demarcate the total Higgs boson mass gradients, including the tree-level plus one/two-loop (as computed by the SuSpect~2.34 codebase), the three-loop plus four-loop contributions, and the flippon contribution. The thin dashed lines represent gradients of tan$\beta$, while the upper and lower exterior boundaries are defined by a top quark mass of $m_t = 173.3 \pm1.1$ GeV. The left edge is marked by the LEP constraints, while the right edge depicts where the Planck relic density can no longer be maintained due to an LSP and light stau mass difference less than the on-shell tau mass. All model space within these boundaries satisfy the Planck relic density constraint $\Omega h^2 = 0.1199 \pm 0.0027$ and the No-Scale requirement $B_{\mu}=0$. Reprinted with permission from~\cite{Li:2013naa}.}
        \label{fig:higgstanb}
\end{figure}

In Refs.~\cite{Li:2011xu,Li:2013naa}, we presented the wedge of No-Scale \fsu5
model space that is consistent with a set of ``bare minimal''
constraints from theory and phenomenology. The constraints included: i) consistency with the dynamically established boundary conditions of No-Scale supergravity (most notably the
imposition of a vanishing $B_{\mu}$ at the final flipped $SU(5)$
GUT unification near $M_{\rm Pl}$, enforced as $\left|B_{\mu}\right(M_{\cal F})| \leq 1$ GeV, about
the size of the EW radiative corrections); ii) radiative electroweak
symmetry breaking; iii) the centrally observed WMAP7 CDM relic density (and
now the Planck relic density $\Omega h^2 = 0.1199 \pm 0.0027$); iv) the world average
top-quark mass $m_t = 173.3 \pm 1.1$~GeV; v) precision
LEP constraints on the light SUSY chargino and neutralino mass
content; and vi) production of a lightest CP-even Higgs boson mass of $m_{h} = 125.5 \pm 1.5$
GeV, accomplished through additional tree level and one-loop contributions to the Higgs boson mass by
the flippon supermultiplets~\cite{Li:2011ab,Li:2012jf,Li:2013naa}, supplementing the Minimal
Supersymmetric Standard Model (MSSM) Higgs boson mass by just the essential additional 3-5 GeV amount
requisite to attain $m_{h} \sim 125$ GeV, while also preserving a testably light SUSY spectrum that does not reintroduce the gauge hierarchy problem via very heavy scalars that SUSY was originally intended to solve in the first place. This two-dimensional parameterization in the
vector-like {\it flippon} super-multiplet mass scale $M_V$
and the universal gaugino boundary mass scale $M_{1/2}$
was excised from a larger four-dimensional hyper-volume
also including the top quark mass $m_t$ and the ratio
$\tan \beta$. Surviving points, each capable
of maintaining the delicate balance required to satisfy 
$B_\mu = 0$ and the CDM relic density observations, were
identified from an intensive numerical scan, employing
MicrOMEGAs~2.1~\cite{Belanger:2008sj} to compute SUSY masses, using a proprietary modification of the
SuSpect~2.34~\cite{Djouadi:2002ze} codebase to run the
{\it flippon}-enhanced RGEs.

The union of all such points was found to consist of
a diagonal wedge ({\it cf.} Ref.~\cite{Li:2011xu,Li:2013naa}) in the $M_{1/2}$-$M_V$
plane, the width of which ({\it i.e.} at large $M_{1/2}$ and
small $M_V$ or vice-versa) is bounded by the central 
experimental range of the top quark mass, and the extent
of which ({\it i.e.} at large $M_{1/2} \sim 1500$~GeV and large $M_V$) is
bounded by CDM constraints and the transition to a charged
stau LSP. This upper region of the model space corresponds to an exponentially elevated {\it flippon}
mass $M_V$, which may extend into the vicinity of 100~TeV.  This
delineation of the bare-minimally constrained \fsu5 parameter
space, including the correlated values of $m_t$, $\tan \beta$ and
the light CP-even Higgs mass for each model point, is depicted in
Figure~\ref{fig:higgstanb}. One obvious concern associated with this circumstance is the appearance
of a new intermediate scale of physics, and a potentially new associated
hierarchy problem.  However, we remark that the vector-like {\it flippon}
multiplets are free to develop their own Dirac mass, and are not in
definite {\it a priori} association with the electroweak scale symmetry
breaking; We shall therefore not divert attention here 
to the mechanism of this mass generation, although plausible
candidates do come to mind.

The advent of substantial LHC collision data in
the SUSY search rapidly eclipsed the tentative low-mass
boundary set by LEP observations.  A substantive correlation
in the \fsu5 mass scale favored by low-statistics excesses
in a wide range of SUSY search channels, particularly lepton-inclusive searches, at both CMS and ATLAS was remarked upon by
our group~\cite{Li:2013hpa} just below $M_{1/2} \sim 800$~GeV.
However, a minority of search channels, particularly lepton-exclusive
squark and gluino searches with jets and missing energy~\cite{ATLAS-CONF-2012-109},
were found to yield limits on $M_{1/2}$ that are inconsistent
with this fit, and that exert some limited tension against the upper
$M_{1/2}$ boundary of the model wedge.  This tension is also
reflected in one generic limit of a multijet plus a single lepton SUSY search from the CMS Collaboration
that places the gluino heavier than about 1.3~TeV~\cite{Chatrchyan:2013iqa}.

\section{No-Scale \bfsu5 at the $\sqrt{s}$ = 14 TeV LHC}

The LHC will be resuming collisions during mid 2015, initiating Run 2 at a center of mass energy of
$\sqrt{s} = 13$~TeV.  It is expected that an increase to the full design energy of 14 TeV
will be undertaken sometime thereafter, similar to the staggered 7/8 TeV operation of Run 2.
This section summarizes the outlook for No-Scale \fsu5 at these higher energies.  It is based 
upon an analysis~\cite{Dutta:2014zua} performed in collaboration with Bhaskar Dutta and Kuver Sinha,
and in consultation with Teruki Kamon, whom we presently thank and acknowledge.

A key driver of the collider phenomenology in No-Scale \fsu5 is the fact that the
light stop $\tilde{t}_1$ is lighter than the gluino,
and by an amount that allows for on-shell decays with unity branching ratio for most of the viable parameter space.
The strongest signal of new physics in a $m(\tilde{q}) > m(\tilde{g}) > m({\tilde{t}}_1)$ type model
is expected in association with extremely long cascade decay chains, featuring a strong
four $W$ plus four $b$ heavy flavor jet component~\cite{CMS:2014dpa}.
Since the $W$ may decay leptonically ($1/3$ for three light generations) or
hadronically ($2/3$ for two light generations times 3 colors), the final state will also be profuse
with leptons and multi-jets.  In order to establish the signal, we
therefore require at least two $b$-jets in all cases,
while recording the net count of jets, leptons, di-leptons, and missing transverse energy $\met$,
expecting $(i)$ that events with fewer leptons should have more jets, and $(ii)$ that
the dominant $\ttbar+ {\rm Jets}$ background may likewise have large jet counts, but should not
generally feature very large $\met$ values. 
For $t\overline{t} + {\rm Jets}$, charge conservation further implies that any dilepton production
must be anti-correlated in sign, whereas the independent leptonic decay
events are uncorrelated in flavor. The SUSY four $W+b$ signal may readily
produce tri-leptons (category III), which are inaccessible, outside of fakes, to
$t\overline{t} + {\rm Jets}$; this category, which necessarily 
includes also a like-sign dilepton, should be intrinsically low background.
Likewise, the orthogonal categorization of precisely two leptons (category II)
with like sign should intrinsically suppress $t\overline{t} + {\rm Jets}$,
with residual fakes, sign-flips, etc., reduced by a requirement on
missing transverse energy $\met$.  The remaining event subdivisions (category I),
i.e. those with $0,1,~{\rm or}~2$ leptons, but no like-sign dilepton,
will rely heavily on the missing energy cut for background reduction,
but may also feature a much stronger net signal count.  Opposite sign
di-tau production, which is important for distinguishing between
thermal and non-thermal dark matter production mechanisms~\cite{Dutta:2014zua}, is a
small subset of this very broad event category.

\begin{figure}[htp]
        \centering
        \includegraphics[width=0.45\textwidth]{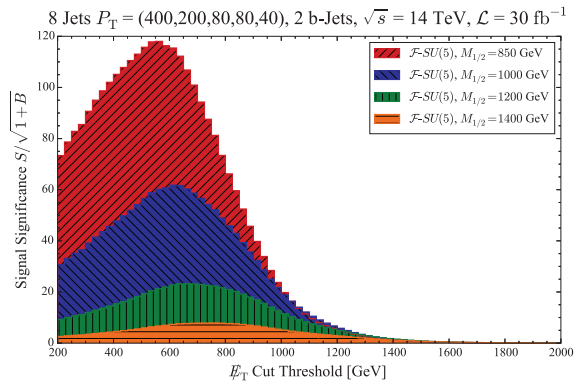}
        \caption{
Events with (0,1,2) leptons and no like-sign dilepton (category I)
are studied at a luminosity of 30 events per femtobarn.
Two heavy-flavor tagged jets with $P_{\rm T} > 80$~GeV are required, and the leading
eight jets (with or without a b-tag) must carry $P_{\rm T} > (400,200,80,80,40)$~GeV.
Signal significance relative to the leading $\ttbar$ background
is evaluated for four signal regions as a function of the $\met$ cut.
	}
        \label{fig:catI}
\end{figure}

\begin{figure}[htp]
        \centering
        \includegraphics[width=0.45\textwidth]{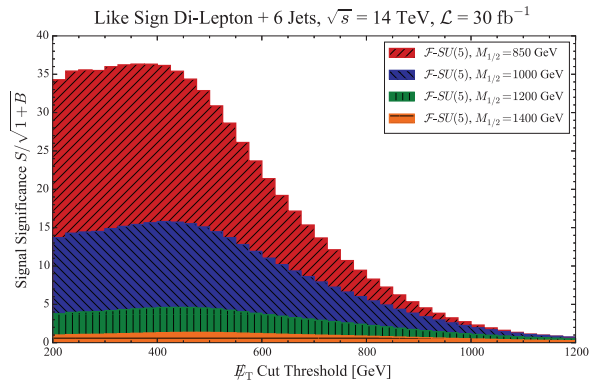}
        \caption{
Events with a like-sign dilepton topology (category II)
are studied at a luminosity of 30 events per femtobarn.
Two heavy-flavor tagged jets with $P_{\rm T} > 80$~GeV are required, and the leading
four jets (with or without a b-tag) must carry $P_{\rm T} > (400,200,80,80)$~GeV.
Two additional jets (for a total of 6) must carry $P_{\rm T} > 40$~GeV.
Signal significance relative to the leading $\ttbar$ background
is evaluated for four signal regions as a function of the $\met$ cut.
	}
        \label{fig:catII}
\end{figure}

\begin{figure}[htp]
        \centering
        \includegraphics[width=0.45\textwidth]{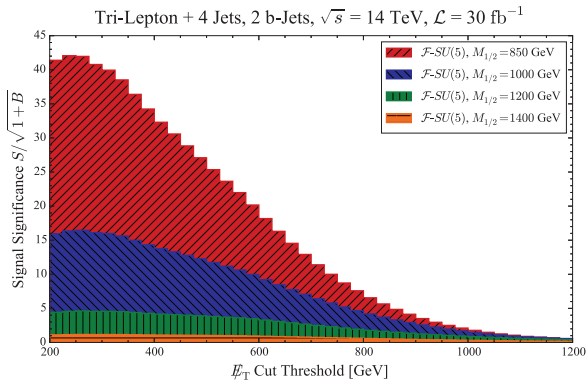}
        \caption{
Events with a trilepton topology (category III)
are studied at a luminosity of 30 events per femtobarn.
Two heavy-flavor tagged jets with $P_{\rm T} > 80$~GeV are required, and the leading
four jets (with or without a b-tag) must carry $P_{\rm T} > (400,200,80,80)$~GeV.
Signal significance relative to the leading $\ttbar$ background
is evaluated for four signal regions as a function of the $\met$ cut.
	}
        \label{fig:catIII}
\end{figure}

The described 2 $b$-jet signal categories (I,II,III) corresponding to prominent
signals associable with models featuring long-chain decay cascades with a light
third generation have been established in Monte Carlo simulation.
Signal and standard model (SM) background,
including parton showering and fast detector simulation, are generated via the standard
{\sc MadGraph5}/{\sc MadEvent}~\cite{Alwall:2011uj}, {\sc Pythia}~\cite{Sjostrand:2006za}, {\sc PGS4}~\cite{PGS4} chain.
Event selection and analysis is performed with {\sc AEACuS 3.6}~\cite{Walker:2012vf,aeacus}.
The intuition that something
like $8,6,4$ hard jets may be respectively expected in the signal for each category
is well confirmed, noting that the latter categories exchange
jet pair production for lepton production in the $W$ decay.  Any jets associated
with a squark to gluino transition (typically a 500 GeV to 750 GeV mass gap)
are expected to be quite hard.  Jets downstream from the stop decay also
receive a substantial boost from the mass differential, and all downstream
jets may inherit large kinematic boosts, even in decays with less phase space.
Requiring $P_{\rm T}^{1,2} > (400,200)$~GeV facilitates very robust tagging
on the leading jet pair, while dampening background (allowing a lower $\met$ floor),
and retaining excellent signal statistics.  Jets 3 and 4 are well resolved at
$P_{\rm T}^{3,4} > 80$~GeV, consistent with the $b$-jet threshold, whereas any jets
required beyond the leading four are better captured with softer threshold
around $P_{\rm T}^{5+} > 40$~GeV.  With these cuts in place, the missing transverse
energy $\met$ threshold may be individually optimized for each category,
as demonstrated in Figures~\ref{fig:catI},\ref{fig:catII},\ref{fig:catIII}.
We will select $\met > (700,500,300)$~GeV, respectively.

The background is found in each case to be extraordinarily well controlled,
with excellent signal retention.  Categories III (tri-leptons) and II
(like-sign di-leptons) appear to be observable up to about $M_{1/2} \sim 1200$~GeV,
while the primary category I (all other events) may be probed beyond $M_{1/2} \sim 1400$~GeV,
encompassing the majority, if not totality, of the \fsu5 model space.  The gluino
masses in these cases are on the order of 1600 and 1900 GeV, respectively. As
demonstrated clearly in Figure~\ref{fig:catI}, the expected SUSY event yield is
a strongly decaying function of $M_{1/2}$, which may be inverted in order to
establish the global model mass scale.  Since the model is dominantly single
parameter, the bulk properties of the spectrum are then fixed, and may be
cross-correlated against alternatively designed event selections for consistency,
such as the di-tau production channel.

\section{Electroweak Fine-Tuning in No-Scale \bfsu5}

We detail in this section the minimization of electroweak fine-tuning in No-Scale \fsu5,  based upon analyses~\cite{Leggett:2014mza,Leggett:2014hha},
which the authors of this work completed in collaboration with Tristan Leggett, whose contribution we acknowledge.

The SUSY framework naturally provides for interplay between quartic and quadratic field strength terms
in the scalar potential of the type essential to spontaneous destabilization of the null vacuum, the former emerging
with dimensionless gauge-squared coupling coefficients from the $D$-term, and the latter with dimensionful mass-squared
coefficients referencing the bilinear Higgs mixing scale $\mu$ from the chiral $F$-term.
Crucially though, this radiative EWSB event, as driven by largeness of the top-quark Yukawa coupling, is not realizable without the supplementary inclusion of soft mass terms $m_{H_{u,d}}$ and the analog $B_\mu$ of $\mu$, which herald first the breaking of SUSY itself.  In a supergravity (SUGRA) context, these terms may be expected to appear in proportion to the order parameter of SUSY breaking in the visible sector, as gravitationally suppressed from higher scale effects in an appropriately configured hidden sector, namely the gravitino mass $M_{3/2}$.  The gravitino mass may itself be exponentially radiatively suppressed relative to the high scale, plausibly and naturally taking a value in the TeV range.  The Giudice-Masiero (GM) mechanism may be invoked to address the parallel ``$\mu$ problem'', suggesting that this SUSY-preserving coupling may likewise be of the same order, and likewise generated as a consequence of SUSY breaking, as evaluated at the high scale.

Minimization of the Higgs scalar potential with respect to the $H_u$ and $H_d$ field directions
yields two conditions on the pair of resulting vacuum expectation values (VEVs) $(v_u,v_d)$.  The overall scale
$(v_u^2+v_d^2)^{1/2}$, in product with the gauge coefficients $(g_{\rm L}^2+{g'}_{\rm Y}^2)^{1/2}/2$, is 
usually traded for the physical $Z$-boson mass $M_Z$, whereas the relative VEV strengths are parameterized
$(\tan \beta \equiv v_u/v_d)$ by an angle $\beta$.  This allows one to solve for $\mu$ and $B_\mu$ at the
electroweak scale in terms of $M_Z$, $\tan \beta$, and the soft masses $m_{H_{u,d}}$.  When addressing the
question of fine tuning, the solution for $\mu^2$ is typically inverted as follows in Eq.~(\ref{eq:ewmin}), and an argument is
made regarding the permissible fraction of cancellation between terms on the right-hand side, whose
individual scales may substantially exceed $M_Z^2$. 
\begin{eqnarray}
\frac{M_Z^2}{2} =
\frac{m_{H_d}^2  -
\tan^2\beta ~m_{H_u}^2}{\tan^2\beta -1} -\mu^2
\label{eq:ewmin}
\end{eqnarray}

\noindent For moderately large $\tan\beta$, Eq.~(\ref{eq:ewmin}) reduces to
\begin{eqnarray}
\frac{M_Z^2}{2} \simeq -m_{H_u}^2 -\mu^2 \;.
\label{eq:EWMINL}
\end{eqnarray}

Several approaches to quantifying the amount of fine tuning implicit in Eq.~(\ref{eq:ewmin}) have been suggested,
one of the oldest being that $\Delta_{\rm EENZ}$~\cite{Ellis:1986yg, Barbieri:1987fn} first prescribed some 30 years ago by
Ellis, Enqvist, Nanopoulos, and Zwirner (EENZ), consisting of the maximal logarithmic $M_Z$ derivative with respect
to all fundamental parameters $\varphi_i$, evaluating at some high unification scale $\Lambda$ as is fitting for gravity-mediated SUSY breaking.
In this treatment, low fine-tuning mandates that heavy mass scales only weakly influence $M_Z$, whereas strongly correlated
scales should be light.
\begin{eqnarray}
\Delta_{\rm EENZ} = {\rm Max}\,\left\{\,
\left|
\frac{\partial\,{\rm ln}(M_Z^n)}{\partial\, {\rm ln}(\varphi_i^n)}
\right|\,
\equiv\,
\left|
\frac{\varphi_i}{M_Z}
\frac{\partial\,M_Z}{\partial\,\varphi_i}
\right|\,
\right\}_{\Lambda}
\label{eq:eenz}
\end{eqnarray}

In the SUGRA context, $M_Z^2$ is generically bound to dimensionful inputs $\varphi_i$ at the high scale $\Lambda$
via a bilinear functional, as shown following.  The parameters $\varphi_i$ may include scalar and gaugino
soft SUSY breaking masses (whether universal or not), the bi- and tri-linear soft terms $B_\mu$ and $A_i$, as well as the $\mu$-term.
The coefficients $C_i$ and $C_{ij}$ are calculable, in principle, under the renormalization group dynamics.
\begin{eqnarray}
M_Z^2 = \sum_i C_i \varphi_i^2(\Lambda) + \sum_{ij} C_{ij} \varphi_i(\Lambda) \varphi_j(\Lambda)
\label{eq:bilinear}
\end{eqnarray}

\noindent Applying the Eq.~(\ref{eq:eenz}) prescription, a typical contribution to the fine tuning takes the subsequent form.
\begin{eqnarray}
\frac{\partial\,{\rm ln}(M_Z^2)}{\partial\, {\rm ln}(\varphi_i)}
= \frac{\varphi_i}{M_Z^2} \times \bigg\{\, 2\,C_i\varphi_i + \sum_{j} C_{ij} \varphi_j \,\bigg\}
\label{eq:eenzsum}
\end{eqnarray}

\noindent
Comparing with Eq.~(\ref{eq:bilinear}), each individual term in the Eq.~(\ref{eq:eenzsum}) sum is observed,
modulo a possible factor of 2, to be simply the ratio of one contribution to the unified $M_Z^2$ mass, divided by $M_Z^2/2$.

The EW fine-tuning was numerically computed for No-Scale \fsu5 according to the Eq.~(\ref{eq:eenz}) prescription
in Ref.~\cite{Leggett:2014mza}, yielding result of ${\cal O}(1)$.  This absence of fine-tuning is equivalent to
a statement that the $Z$-boson mass $M_Z$ can be predicted in \fsu5 as a parameterized function of $M_{1/2}$; clarifying
and rationalizing this intuition in a more quantitative manner is a key intention of the present section.
First, we define a dimensionless ratio $c$ of the supersymmetric Higgs mixing parameter $\mu$ at the
unification scale $M_{\cal F}$ with the gaugino mass $M_{1/2}$.
\begin{eqnarray}
c = \frac{\mu(M_{\cal F})}{M_{1/2}}
\label{eq:c}
\end{eqnarray}

\noindent This parameter $c$ is a fixed constant if the $\mu$ term is generated via the Giudice-Masiero
mechanism~\cite{Giudice:1988yz}, which can, in principle, be computed from string theory. We adopt an {\it ans\"atz}
\begin{eqnarray}
M_Z^2 = f_1 + f_2\,M_{1/2} + f_3\,M_{1/2}^2
\label{eq:mzquad}
\end{eqnarray}

\noindent consistent with Eq.~(\ref{eq:bilinear}), where the undetermined coefficients $f_i$ represent implicit functions
of dimensionless quantities including $c$ and $\lambda$.  Some evidence suggests that the dimensionful coefficients
$f_1$ and $f_2$ may additionally be sensitive to $B_\mu$, particularly to any potential deviations from the null
No-Scale boundary value.  If $(f_1 \lll M_{1/2}^2)$ and $(f_2 \ll M_{1/2})$, then a linearized approximation of the prior is applicable:
\begin{eqnarray}
M_Z = f_a + f_b\, M_{1/2}~~.
\label{eq:mzlinear}
\end{eqnarray}

The form of  Eq.~(\ref{eq:mzquad}) must now be verified with explicit RGE calculations.   This is accomplished
via a numerical sampling, wherein the $Z$-boson mass is floated within
$20 \leq M_Z \leq 500$ GeV, and the top quark mass (equivalently its Yukawa coupling) within $125 \leq m_t \leq 225$ GeV.
The region scanned for the gaugino mass boundary is within $100 \leq M_{1/2} \leq 1500$ GeV.
In order to truncate the scanning dimension, $M_V$ and $\tan \beta$ are explicitly parameterized functions of $M_{1/2}$
(consistent with the prior description) such that the physical region of the model
space corresponding to $M_Z = 91.2$ GeV and $m_t = 174.3$, along with a valid thermal relic density, is continuously intersected\footnote{
The No-Scale \fsu5 model space favors a top quark mass of $m_t = 174.3 - 174.4$
GeV in order to compute a Higgs boson mass of $m_h \sim 125$
GeV~\cite{:2012gk,:2012gu,Aaltonen:2012qt}.  The central world average 
top quark mass has recently ticked upward (along with an increase in precision) to $m_t = 174.34$ GeV~\cite{TEWG:2014cka},
affirming this preference.}; this may be considered equivalent to fixing the top quark Yukawa coupling (and associated higher-order feedback) within just this subordinate parameterization.
The range of the ratio $c$ from Eq.~(\ref{eq:c}) is an output of this analysis, which is run from the EWSB scale up to $M_{\cal F}$ under the RGEs.

\begin{figure}[htp]
        \centering
        \includegraphics[width=0.45\textwidth]{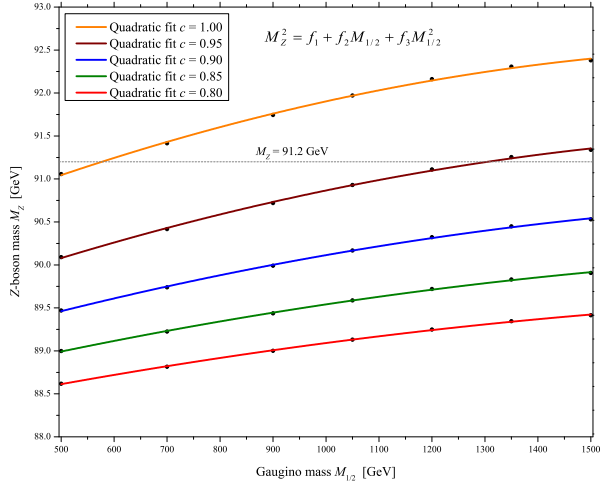}
        \caption{Simple quadratic fits for $M_Z^2$ as a function of $M_{1/2}^2$. Five different cases of $c$ are shown.
The curves are only comprised of points with a vanishing $B_{\mu}$ parameter at the $M_{\cal F}$ unification scale. Reproduced with thanks from~\cite{Leggett:2014hha}.}
        \label{fig:quadratic}
\end{figure}

\begin{figure}[htp]
        \centering
        \includegraphics[width=0.45\textwidth]{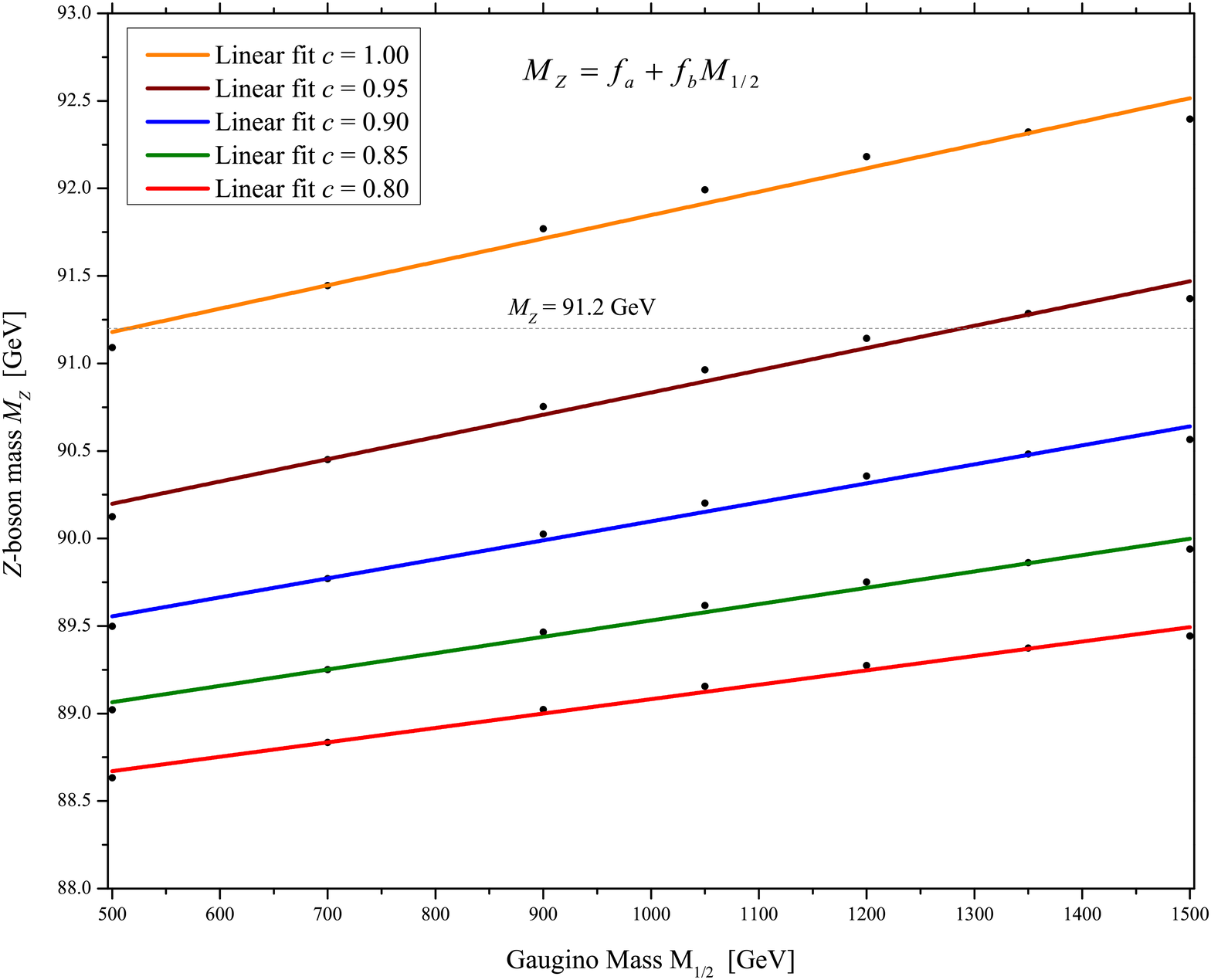}
        \caption{Linear fits for $M_Z$ as a function of $M_{1/2}$. Five different cases of $c$ are shown.
The curves are only comprised of points with a vanishing $B_{\mu}$ parameter at the $M_{\cal F}$ unification scale. Reproduced with thanks from~\cite{Leggett:2014hha}.}
        \label{fig:linear}
\end{figure}

\begin{figure}[htp]
        \centering
        \includegraphics[width=0.45\textwidth]{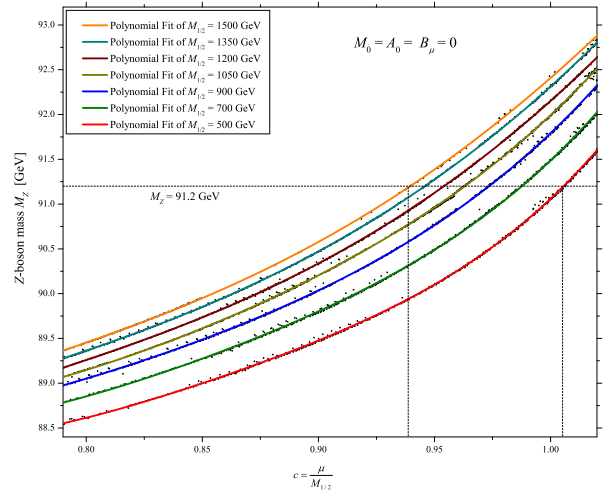}
        \caption{The $Z$-boson mass is shown as a function of the dimensionless parameter $c$ for seven
different values of $M_{1/2}$. The black points are the results of the RGE calculations, while the
curves are polynomial fits. The curves are only comprised of points with a vanishing $B_{\mu}$ parameter at the $M_{\cal F}$ unification scale. Reproduced with thanks from~\cite{Leggett:2014hha}.}
        \label{fig:seven_curves_c}
\end{figure}

The dimensionless parameter $c$ is expected to be a fixed constant if the $\mu$ term is generated by the
GM mechanism. Via the numerical procedure detailed in Ref.~\cite{Leggett:2014hha}, 
this is made explicit for $c = 0.80, 0.85, 0.90, 0.95, 1.00$ in Figure~\ref{fig:quadratic}, where each curve is well fit by a quadratic in the form of Eq.~(\ref{eq:mzquad}).  Figure~\ref{fig:linear} demonstrates a fit against the linear approximation in Eq.~(\ref{eq:mzlinear}).  As the $c$ parameter decreases, Figure~\ref{fig:linear} illustrates that the linear fit approaches the precision of the quadratic fit.  The dimensionful intercept $f_a$ is a function of $c$, but is observed generically to take a value in the vicinity of $89$~GeV.  As seen in Figure~\ref{fig:seven_curves_c}, larger values of $M_{1/2}$ correlate with smaller values of $c$ at fixed $Z$-boson mass; it is
the region $M_{1/2} \gtrsim 900$ GeV that remains viable for probing 
a prospective SUSY signal at the 13--14 TeV LHC in 2015--16.

\begin{figure}[htp]
        \centering
        \includegraphics[width=0.45\textwidth]{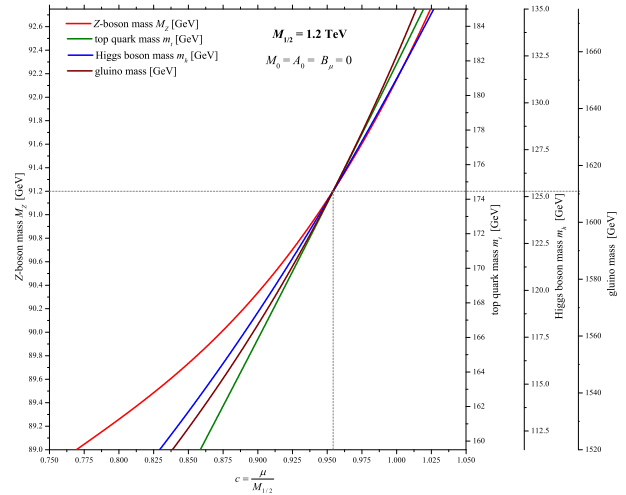}
        \caption{Depiction of the correlation between the $Z$-boson mass $M_Z$, top quark mass $m_t$, Higgs
boson mass $m_h$, and gluino mass $m_{\widetilde{g}}$, as a function of $c$, for $M_{1/2} = 1.2$ TeV.
All other SUSY particles can be expressed similarly. The curves are only comprised of points with a
vanishing $B_{\mu}$ parameter at the $M_{\cal F}$ unification scale. All other $M_{1/2}$ produce comparable correlations. Reproduced with thanks from~\cite{Leggett:2014hha}.}
        \label{fig:spectrum}
\end{figure}

The No-Scale SUGRA constraint on the $B_{\mu}$ parameter naturally parameterizes all the particle and
sparticle masses as a function of the dimensionless parameter $c$ of Eq.~(\ref{eq:c}). This is clearly
shown in Figure~\ref{fig:spectrum} for the $Z$-boson mass $M_Z$, top quark mass $m_t$, Higgs boson mass
$m_h$, and gluino mass $m_{\widetilde{g}}$. We use the gluino mass as an example, though the entire SUSY
spectrum can also thusly be parameterized as a function of $c$ via the $B_{\mu} = 0$ condition. The point
chosen in Figure~\ref{fig:spectrum} to exhibit the correlation between the particle and sparticle masses
is $M_{1/2} = 1200$ GeV.  Table~\ref{tab:results} itemizes numerical results from the RGE
calculations for  $M_{1/2} = 1200$ GeV. The Higgs boson mass $m_h$ in
Table~\ref{tab:results} includes both the tree level+1-loop+2-loop+3-loop+4-loop
contributions~\cite{Li:2013naa} and the additional flippon contribution~\cite{Li:2011ab}.
Sensitivity is observed to fluctuation of the VEV scale with $M_Z$. To be concrete, we present
a benchmark point with $M_{1/2}=990$~GeV in Table~\ref{bench1}.
This example is in the stau-neutralino coannihilation region, with thermal Bino dark matter providing 
the observed relic density. The selected mass range is in the vicinity of the exclusion boundary established 
data from the 7--8~TeV LHC runs;
commencement of collisions near the 13--14~TeV design energy will actively probe the \fsu5 construction 
at scales above $M_{1/2} = 1$~TeV.

\begin{table}[htbp!]
\centering
\caption{Results of RGE calculations for $M_{1/2}$ = 1.2~TeV. These are only points with a
vanishing $B_{\mu}$ parameter at the $M_{\cal F}$ unification scale. The entries highlighted in red are those that
compute the observed experimental measurements for $M_Z$, $m_t$, and $m_h$. Other values of $M_{1/2}$ show similar results~\cite{Leggett:2014hha}.}
\begin{tabular}{|c|c|c|c|c|}\cline{1-5}
\multicolumn{5}{|c|}{$M_{1/2} = 1200~ {\rm GeV}$} \\ \hline
$c$&$M_Z$&$m_t$&$m_h$&$m_{\widetilde{g}}$ \\ \hline
$	0.913	$&$	90.525	$&$	167.96	$&$	120.31	$&$	1574	$ \\ \hline
$	0.923	$&$	90.672	$&$	169.49	$&$	121.48	$&$	1583	$ \\ \hline
$	0.933	$&$	90.832	$&$	171.06	$&$	122.66	$&$	1591	$ \\ \hline
$	0.943	$&$	90.999	$&$	172.63	$&$	123.88	$&$	1601	$ \\ \hline
${\color{red}	0.953}	$&${\color{red}	91.180}	$&${\color{red}	174.24}	$&${\color{red}	125.09}	$&${\color{red}	1610}	$ \\ \hline
$	0.963	$&$	91.371	$&$	175.87	$&$	126.41	$&$	1619	$ \\ \hline
$	0.970	$&$	91.502	$&$	176.94	$&$	127.26	$&$	1626	$ \\ \hline
$	0.983	$&$	91.780	$&$	179.11	$&$	129.04	$&$	1640	$ \\ \hline \hline
\end{tabular}
\label{tab:results}
\end{table}

\begin{table}[ht]
  \small
    \caption{Spectrum (in GeV) for $M_{1/2} = 990$~GeV, $M_{V} = 8044$~GeV, $m_{t} = 174.4$~GeV, and $\tan \beta$ = 23.3.
	Here, $\Omega_{\rm CDM} h^2$ = 0.1197, the stau-LSP mass gap is $\Delta M = 6.4$~GeV, and the lightest neutralino is greater than 99\% Bino.
        For other values of $M_{1/2}$, revisions to the complete SUSY spectrum may be very well approximated by a simple proportional rescaling.  $\Delta M$
        may be increased by slightly lowering $\tan \beta$, with minimal additional effect on the spectrum overall.}
\hspace{-5pt}\begin{minipage}{\linewidth}
\begin{tabular}{|c|c||c|c||c|c||c|c||c|c||c|c|} \hline
     $\widetilde{\chi}_{1}^{0}$ & $213$ & $\widetilde{\chi}_{1}^{\pm}$ & $449$ & $\widetilde{e}_{R}$ &    $366$ & $\widetilde{t}_{1}$ & $1104$ & $\widetilde{u}_{R}$ & $1824$ & $m_{h}$ &         $125.1$\\ \hline
     $\widetilde{\chi}_{2}^{0}$ & $449$ & $\widetilde{\chi}_{2}^{\pm}$ & $1463$ & $\widetilde{e}_{L}$ &   $989$ & $\widetilde{t}_{2}$ & $1672$ & $\widetilde{u}_{L}$ & $1985$ & $m_{A,H}$ &       $1590$\\ \hline
     $\widetilde{\chi}_{3}^{0}$ & $1461$ & $\widetilde{\nu}_{e/\mu}$ &   $986$ & $\widetilde{\tau}_{1}$ & $220$ & $\widetilde{b}_{1}$ & $1650$ & $\widetilde{d}_{R}$ & $1887$ & $m_{H^{\pm}}$ &   $1592$\\ \hline
     $\widetilde{\chi}_{4}^{0}$ & $1463$ & $\widetilde{\nu}_{\tau}$ &    $958$ & $\widetilde{\tau}_{2}$ & $964$ & $\widetilde{b}_{2}$ & $1789$ & $\widetilde{d}_{L}$ & $1986$ & $\widetilde{g}$ & $1328$\\ \hline
	\end{tabular}
\end{minipage}
\label{bench1}
\end{table}

\begin{figure}[htp]
        \centering
        \includegraphics[width=0.45\textwidth]{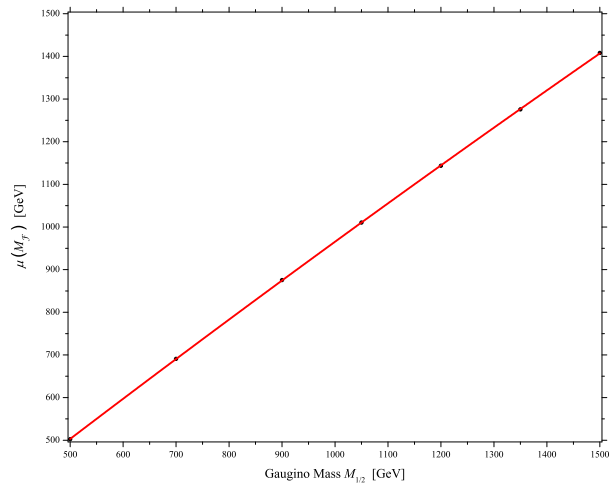}
        \caption{Linear relationship between the $\mu$ term at the $M_{\cal F}$ unification scale and $M_{1/2}$ for $M_Z =91.2$ GeV. Reproduced with thanks from~\cite{Leggett:2014hha}.}
        \label{fig:mu}
\end{figure}

The relationship between the $\mu$ term and $M_{1/2}$ at the $M_{\cal F}$ unification scale is linear
for fixed $M_Z$, with a slope given by the ratio $c$ from Eq.~(\ref{eq:c}).  This is expanded in 
Figure~\ref{fig:mu} for $M_Z = 91.2$~GeV.

Having established a (family in $c$ of) quadratic expression(s) for $M_Z^2$ in the Eq.~(\ref{eq:bilinear}) form, the $Z$-boson mass is extracted by reference only to $M_{1/2}$ and $c$ at the high scale $\Lambda$, and fine tuning may be evaluated.  Adopting the linear Eq.~(\ref{eq:mzlinear}) form, we first consider tuning with respect to $M_{1/2}$ at fixed $c$,
as prescribed by Eq.~(\ref{eq:eenz}).
\begin{eqnarray}
\Delta_{M_{1/2}}^{\rm \Lambda} &=& \left| \frac{\partial{\rm ln}(M_Z)}{\partial {\rm ln}(M_{1/2})} \right|
= \left| \frac{M_{1/2}}{M_Z} \frac{\partial M_Z}{\partial M_{1/2}} \right|  \\ \nonumber 
&=& \frac{1}{M_Z} \left( \frac{M_Z - f_a}{f_b} \right) f_b = 1 - \frac{f_a}{M_Z} \simeq 1 - \frac{89}{M_Z} \\ \nonumber
\label{eq:fnpar}
\end{eqnarray}

\noindent Curiously, this expression evaluates very close to zero.  It would appear this result is a consequence
of the fact that the physical $Z$-boson mass can in fact be stably realized for a large continuum of $M_{1/2}$ values,
at the expense of variation in the ratio $c$.  It may be better understood by attending in turn to the parallel functional dependence on the dimensionless $c$ parameter itself.  We have
\begin{eqnarray}
\Delta_{c}^{\rm \Lambda} &=& \left| \frac{\partial{\rm ln}(M_Z)}{\partial {\rm ln}(c)} \right|
= \left| \frac{c}{M_Z} \frac{\partial M_Z}{\partial c} \right| \\ \nonumber
&=& \left| \frac{c}{M_Z}  \left( \frac{\partial f_a}{\partial c} + \frac{\partial f_b}{\partial c} M_{1/2} \right) \right| \sim c \simeq 1~,~\ \\ \nonumber
&&
\label{eq:fnc}
\end{eqnarray}

\noindent using the numerical observation \mbox{$\frac{\partial f_a}{\partial c} + \frac{\partial f_b}{\partial c} M_{1/2} \sim M_Z$}. Therefore, stipulating the adopted high-scale context, we suggest that the more natural fine-tuning measure for No-Scale \fsu5 may be $\Delta_{\rm EENZ} \sim 1$.

\section{No-Scale Supergravity and the Starobinsky Model}

This section encapsulates the potential fundamental connection between No-Scale SUGRA and the Starobinsky model of cosmological inflation. It is based upon analyses~\cite{Ellis:2013xoa,Ellis:2013nxa,Ellis:2013nka} D.V.N. performed in collaboration with John Ellis and Keith Olive, whom we presently thank and acknowledge.

Recently, an added phenomenological boost has been given to No-Scale Supergravities by detailed
measurement of the Cosmic Microwave Background (CMB) perturbations (the structural seeds of galactic
supercluster formation residually imprinted upon the faint afterglow of the big bang) from the Planck~\cite{Planck:2015xua,Ade:2015lrj} satellite.  This experiment verified a highly statistically significant tilt $n_s < 1$ in the spectrum of
scalar perturbations, and set stronger upper limits on the ratio $r$ of tensor (directional) to
scalar (isotropic) perturbations.  These measurements, particularly of $n_s$, place many leading
models of cosmic inflation in jeopardy.
For example, single-field models with a monomial potential $\phi^n: n \ge 2$
are now disfavored at the $\sim 95$\% CL in the case of $\phi^2$ models, and at higher CLs for
models with $n > 2$. This has revived interest in non-monomial single-field potentials, such as
that found in the minimal Wess-Zumino model~\cite{Croon:2013ana}~\footnote{Models with similar potentials
were proposed  long ago~\cite{Linde:1984cd,Linde1,Albrecht:1984qt} and more recently in~\cite{Kallosh:2007wm}: see~\cite{Olive:1989nu} for a review.}.

A curious scenario suggested by Starobinsky~\cite{Starobinsky:1980te}
in 1980 is known~\cite{Mukhanov:1981xt} to match the CMB data effortlessly,
yielding a value of $n_s \sim 0.96$ that is in perfect accord with experiment,
and a value of $r \sim 0.004$ that is comfortably consistent with the Planck upper limit~\cite{Planck:2015xua,Ade:2015lrj}.
This model is a rather ad-hoc modification of Einstein's description of gravity, which combines a quadratic power of the
Ricci scalar with the standard linear term.  At face value, this model is rather difficult to
take seriously, but there is substantial enthusiasm for the observation that this esoteric model is
in fact conformally equivalent to the low energy limit of No-Scale supergravity with a non-minimal $N_C \ge 2$
scalar sector~\cite{Ellis:2013xoa,Ellis:2013nxa}.  To be specific, the algebraic equations of motion corresponding
to a scalar field $\Phi$ with a quadratic potential that couples to a conventional Einstein term
may be freely substituted back into the action, resulting in the phenomenologically favorable
quadratic power of the scalar curvature~\cite{Stelle:1977ry,Whitt:1984pd}.

In considering the fundamental problem of how cosmological inflation fits into particle physics, a
point of view has been taken~\cite{Ellis:2013xoa,Ellis:2013nxa,Ellis:2013nka} that this union cries out for supersymmetry~\cite{Ellis:1982ed,Ellis:1982dg,Ellis:1982ws}, in
the sense that it requires an energy scale hierarchically smaller than the Planck
scale, thanks to either a mass parameter being $\ll M_P$ and/or a scalar self-coupling
being $\ll {\cal O}(1)$.  Since cosmology necessarily involves consideration of gravity,
it is natural to consider inflation in the context of local supersymmetry, i.e., supergravity~\cite{Freedman:1976xh,Deser:1976eh},
which points in turn to the superstring as a sole contender for the consistent master embedding of quantum gravity.
This preference is complicated, however, by the fact that a generic supergravity theory
has supersymmetry-breaking scalar masses of the same order as the gravitino mass, 
giving rise to the so-called $\eta$ problem~\cite{Copeland:1994vg,Stewart:1994ts} (Also see, for example, Refs.~\cite{Lyth:1998xn,Martin:2013tda}), where the large vacuum energy density
during inflation leads to masses for all scalars of order the Hubble parameter \cite{Goncharov:1984qm}.
While inflationary models in simple supergravity can be constructed 
to avoid the $\eta$ problem \cite{Nanopoulos:1982bv,Holman:1984yj}, these models rely on a seemingly
accidental cancellation, invoking extraneous fine tuning in the inflaton mass \cite{Linde:2007jn}.

For this reason, No-Scale supergravity has long been advocated~\cite{Cremmer:1983bf,Ellis:1983sf,Ellis:1983ei,Ellis:1984bm,Lahanas:1986uc} as the unique natural framework for
constructing models of inflation~\cite{Ellis:1984bf,Goncharov:1985ka,Binetruy:1987xj,Murayama:1993xu,Antusch:2009ty}, representing a low energy
limit of the superstring.  Moreover, this construction yields very successful low energy phenomenology, while invoking a bare
minimum (one or zero) of freely adjustable parameters.  These proposals have recently been reinvigorated in light of the
Planck data, constructing an SU(2,1)/SU(2) $\times$ U(1) No-Scale version of the minimal 
Wess-Zumino model~\cite{Ellis:2013xoa,Ellis:2013nxa,Ellis:2013nka}~\footnote{For an alternative supergravity incarnation of the
Wess-Zumino inflationary model, see~\cite{Nakayama:2013jka}.}.
It was shown that this NSWZ model is consistent with the Planck data for a
range of parameters that includes a special case in which it reproduces {\it exactly}
the effective potential and hence the successful predictions of the Starobinsky $R + R^2$ model~\cite{Ellis:2013xoa,Ellis:2013nxa,Ellis:2013nka}.

Starobinsky considered in 1980~\cite{Starobinsky:1980te} a generalization of the Einstein-Hilbert action to contain 
an $R^2$ contribution,
where $R$ is the scalar curvature:
\begin{equation}
S=\frac{1}{2} \int d^4x \sqrt{-g} (R+\alpha R^2) \, ,
\label{Staro}
\end{equation}
where $\alpha = 1/6M^2$, and $M \ll M_P$ is some mass scale. As was shown by Stelle in 1978~\cite{Stelle:1977ry} and by Whitt in 1984~\cite{Whitt:1984pd}, 
the theory (\ref{Staro}) is conformally equivalent to a theory combining canonical gravity with a scalar field $\varphi$,
described by
\begin{equation}
S=\frac{1}{2} \int d^4x \sqrt{-g} \left[(1 + 2\alpha \varphi) R - \alpha \varphi^2 \right] \, ,
\label{Whitt}
\end{equation}
as can be seen trivially using the Lagrange equation for $\varphi$ in (\ref{Whitt}).
Making the Weyl rescaling $\tilde{g}_{\mu\nu} = (1 + 2 \alpha \varphi) g_{\mu\nu}$,
equation (\ref{Whitt}) takes the form
\begin{equation}
S=\frac{1}{2} \int d^4x \sqrt{-g} \left[ R + \frac{6 \alpha^2 \partial^\mu \varphi \partial_\mu \varphi}
{(1 + 2 \alpha \varphi)^2} - \frac{\alpha \varphi^2}{(1 + 2 \alpha \varphi)^2} \right] \, .
\label{Cecotti4}
\end{equation}
Making now the field redefinition $\varphi^\prime = \sqrt{\frac{3}{2}} \ln \left( 1+ \frac{\varphi}{3 M^2} \right)$, 
one obtains a scalar-field action with a canonical kinetic term:
\begin{equation}
S=\frac{1}{2} \int d^4x \sqrt{-\tilde{g}} \left[\tilde{R} + (\partial_\mu \varphi^\prime)^2 - \frac{3}{2} M^2 (1- e^{-\sqrt{2/3}\varphi^\prime})^2 \right] \, ,
\end{equation}
in which the scalar potential takes the form
\begin{equation} 
V =  \frac{3}{4} M^2 (1- e^{-\sqrt{2/3}\varphi^\prime})^2 \, .
\label{r2pot}
\end{equation}
The spectrum of cosmological density perturbations found by using (\ref{Staro})
for inflation were calculated by Mukhanov and Chibisov in 1981~\cite{Mukhanov:1981xt} and by
Starobinsky in 1983~\cite{Starobinsky:1983zz}. The current data on cosmic microwave
background (CMB) fluctuations, in particular those from the Planck satellite~\cite{Planck:2015xua,Ade:2015lrj},
are in excellent agreement with the predictions of this $R + R^2$ model.

Some general features of the effective low-energy theory derived from a generic supergravity theory are recalled from Refs.~\cite{Ellis:2013xoa,Ellis:2013nxa,Ellis:2013nka}. Neglecting gauge interactions, which are inessential for our purposes, any such theory
is characterized by a K\"ahler potential $K(\phi_i,  \phi^*_j)$, which is a hermitian function
of the chiral fields $\phi_i$ and their conjugates $\phi^*_j$, and a superpotential $W(\phi_i)$,
which is a holomorphic function of the $\phi_i$, via the combination
$G \equiv K + \ln W + \ln W^*$.  The effective field theory contains a generalized kinetic energy term
\begin{equation}
{\cal L}_{KE} \; = \; K^{ij^*} \partial_\mu \phi_i \partial \phi^*_j \, ,
\label{LK}
\end{equation}
where the K\"ahler metric $K^{ij^*} \equiv \partial^2 K / \partial \phi_i \partial \phi^*_{j}$, and the
effective scalar potential is
\begin{equation}
V \; = \; e^G \left[ \frac{\partial G}{\partial  \phi_i} K_{ij^*}  \frac{\partial G}{\partial  \phi^*_j} - 3 \right] \, ,
\label{effpot}
\end{equation}
where $K_{ij^*}$ is the inverse of the K\"ahler metric. Inserting into Eq.~(\ref{effpot}) the K\"ahler potential of Eq.~(\ref{NS-Kahler}), for N=2 and using the Wess-Zumino Superpotential 

\begin{equation}
W=\frac{\widehat{\mu}}{2} \Phi^2 - \frac{\lambda}{3} \Phi^3,
\label{superpotential}
\end{equation}
with $\widehat{\mu} = \mu (c/3)^{1/2}$, $c = 2<{\rm Re}~T>$, and $\lambda = \mu/3$,
we get the potential for the real part of the inflaton (see Ref.~\cite{Ellis:2013xoa} for details):      

\begin{equation}
V = \mu^2 e^{-\sqrt{2/3x}} {\rm sinh}^2(x/\sqrt{6}).
\label{inflatonpotential}
\end{equation}
Clearly, the Starobinsky potential of Eq.~(\ref{r2pot}) is identical with the No-Scale WZ potential of Eq.~(\ref{superpotential})!

\section{Conclusions}

The No-Scale \fsu5 model has exhibited compatibility with the dynamically established boundary conditions of No-Scale supergravity, radiative electroweak symmetry breaking, the centrally observed Planck cold dark matter relic density, the world average top-quark mass, precision LEP constraints on the light SUSY chargino and neutralino mass
content, and production of a $125.5 \pm 1.5$ GeV lightest CP-even Higgs boson mass. We considered here the experimental prospects of the No-Scale \fsu5 model at the 13--14~TeV LHC Run 2, the distinctive proportional dependence of all model scales upon the unified gaugino mass $M_{1/2}$ which minimizes electroweak fine-tuning and allows the $Z$-boson mass to be expressed as a function of $M_{1/2}$, and the empirical correlation between the Starobinsky model of cosmological inflation and No-Scale SUGRA.

As the clock ticks closer to the launching of the 13--14~TeV LHC era, we pause to assess the profound implications of a No-Scale \fsu5 high-energy framework beyond those solely arising from supersymmetry. Consider the deep significance of revealing the stringy origins of the macrocosm, a ubiquitous landscape of string vacua, a 4-dimensional universe fomented by intersecting D-branes, a cosmos pervasive with higher dimensional spaces, grand unification, a quantum theory of gravity, the essence of the dark matter scaffolding gravitationally tethering our galactic structure, an early universe guided by flippons, and a multiverse shaped by No-Scale Supergravity. While the mainstream eye remains fixed upon the primary LHC Run 2 target of supersymmetry, we should not overlook the decidedly more substantial aforementioned repercussions of acquiring a supersymmetry signal consistent with No-Scale \fsu5.


\section{Acknowledgments}

This research was supported in part by the DOE grant DE-FG02-13ER42020 (DVN), the Natural Science Foundation of China under grant numbers 11135003, 11275246, and 11475238, the National Basic Research Program of China (973 Program) under grant number 2010CB833000 (TL), and by the SHSU Enhancement Research Grant 2014/2015 (JWW).
We also thank Sam Houston State University for providing high performance computing resources.


\bibliography{bibliography}

\end{document}